\documentclass[12pt]{article}
\usepackage{color}
\usepackage{amsmath,amssymb}
\usepackage[dvipdfm]{graphicx}
\usepackage[dvipdfm]{graphicx}
\usepackage{cancel}

\usepackage[hypertex]{hyperref}

\begin{document}
\baselineskip=16pt
\begin{titlepage}
\begin{flushright}
{\small OU-HET 705/2011}\\
\end{flushright}
\vspace*{1.2cm}
\begin{center}

{\Large\bf 
$\nu$-Two Higgs Doublet Model and its 
 Collider Phenomenology
} 
\lineskip .75em
\vskip 1.5cm

\normalsize
$^1${\large Naoyuki Haba}, 
and
$^2${\large Koji Tsumura}

\vspace{1cm}

$^1${\it Department of Physics, 
 Osaka University, Toyonaka, Osaka 560-0043, 
 Japan} \\

$^2${\it Department of Physics, National Taiwan University, No. 1,
 Section 4, Roosevelt Rd., Taipei, Taiwan}
\vspace*{10mm}

{\bf Abstract}\\[5mm]
{\parbox{13cm}{\hspace{5mm}

Smallness of neutrino masses can be explained 
 by introducing a tiny vacuum expectation value of 
 an extra-Higgs doublet which couples to 
 right-handed neutrinos ($N_R$). 
This situation is naturally realized 
 in $\nu$-Two Higgs Doublet Model ($\nu$THDM), where  
 a TeV-scale seesaw mechanism can work well. 
We investigate observable phenomenology  
 of $\nu$THDM at LHC and ILC experiments. 
Charged Higgs boson ($H^\pm$)
 in $\nu$THDM 
 is almost originated from the
 extra-Higgs doublet and its coupling strength to neutrinos
 are not small. 
Then this model 
 induces 
 rich phenomenology at the LHC,  
 for example, when $m_{H^\pm}^{} < M_N$, 
 observable charged tracks can be induced from  
 long lived charged Higgs. 
On the other hand, 
 when $m_{H^\pm}^{} > M_N$, 
 right-handed neutrinos can be long-lived, and  
 secondary vertices may be tagged at the LHC. 
The $\nu$THDM also predicts observable 
 lepton number violating process at the ILC.
}}

\end{center}

\end{titlepage}

\section{Introduction}

Recent progress in neutrino oscillation experiments 
 gradually reveal a 
 structure of 
 lepton sector\cite{Strumia:2006db, analyses}.  
However, from a theoretical point of view, 
 smallness of neutrino masses is still a mystery 
 in the standard model (SM), and 
 it is one of the
 most important clues to find new physics beyond the SM. 
It is known that the seesaw mechanism naturally realizes tiny  
 masses of active neutrinos 
 through heavy particles coupled with 
 left-handed neutrinos\cite{Type1seesaw}. 
However, 
 those heavy particles are almost decoupled in the low-energy effective
 theory, few observations may be expected in collider experiments. 
Then, a possibility have been discussed to reduce
 seesaw scales to be TeV~\cite{TeVseesaw,Haba:2009sd}, 
 where effects of TeV scale right-handed neutrinos might be 
 observed in collider 
 experiments such as Large Hadron Collider (LHC) and
 International Linear Collider (ILC). 
However, they must introduce fine-tunings in order to obtain 
 both tiny neutrino masses
 and a detectable left-right neutrino mixing 
 through which they insist experimental evidences can be discovered\cite{XGHe}.
Other right-handed neutrino production processes in extended models
 by e.g.,
 $Z'$ exchange~\cite{Zprime, {Ref:B-L}}
 or Higgs/Higgsino decay~\cite{CerdenoSeto}
 have been also pointed out. 

Here let us take 
 a different approach. 
We remind that
 Dirac masses of fermions are proportional to their Yukawa couplings
 as well as a vacuum expectation value (VEV) of the relevant Higgs field.
Hence, the smallness of masses might be due to not a small Yukawa coupling
 but a small VEV of the Higgs field.
Such a situation is indeed realized in some two-Higgs-doublet model (THDM). 
For example, 
 in Type-II THDM, 
 the mass hierarchy between up-type quark and down-type quark 
 can be explained by the ratio $(\tan\beta)$ of VEVs, and 
 when $\tan\beta \sim 40$,
 Yukawa couplings of top- and bottom-quark are the 
 same scale of order of unity~\cite{Hashimoto:2004xp}.
Similarly, there is a possibility that 
 smallness of neutrino masses comparing to those of quarks and
 charged leptons is originating from an extra Higgs
 doublet with the tiny VEV.  
This idea is that 
 neutrino masses are much smaller than other fermions 
 because the origin of them comes from different VEV of different
 Higgs doublet, and then 
 we do not need  
 extremely tiny neutrino Yukawa coupling constants. 
This kind of model has been considered 
 in Refs.~\cite{Ma,Nandi,Ma:2006km,Davidson:2009ha,Logan:2010ag,HabaHirotsu}. 
We call the model 
 $\nu$THDM in this paper, where 
 TeV-scale seesaw can work, and 
 investigate detective 
 possibilities of the model in collider experiments. 
Especially, in models in Refs.\cite{Ma, HabaHirotsu}, 
 tiny Majorana neutrino masses are obtained  
 through a TeV scale Type-I seesaw mechanism 
 without requiring tiny Yukawa couplings. 
The fact that neutrino Yukawa couplings in the $\nu$THDM 
 are not so small, which also makes  
 low energy thermal 
 leptogenesis work\cite{Haba:2011ra}.

As for 
 extending a Higgs sector, 
 there are 
 constraints in general, for example, 
 consistency of 
 electroweak precision data\footnote{
It is pointed out that the second Higgs doublet
 heavier than SM-like Higgs boson 
 can be consistent with precision electroweak data\cite{Barbieri:2006dq}. 
} 
 and 
 absence of large flavor changing neutral
 currents (FCNCs)\cite{2hdm, {Ref:THDM}}.  
In this model, 
 both two constraints are 
 satisfied 
 since the extra scalar doublet only has a Yukawa interaction 
 with lepton doublets and right-handed neutrinos\footnote{
This is a kind of a ``leptophilic Higgs boson'' which could explain 
 PAMERA and ATIC results\cite{Goh:2009wg}. 
}, and 
 their masses are heavy 
 enough to suppress FCNCs
 although its VEV is of order $0.1$ MeV. 
Decay processes of charged Higgs boson can induce 
 important phenomenology in our model,  
 since 
 it is almost originated from the
 extra-Higgs doublet and its couplings to neutrinos
 are not small. 
Note that the decay of the charged Higgs boson to quarks 
 are strongly suppressed due to absence of 
 direct interactions among them, which 
 is a large difference from other THDMs.  
When $m_{H^\pm}^{} < M_N$, 
 charged tracks of long lived charged Higgs boson
 can be found at the LHC. 
On the other hand, 
 when $m_{H^\pm}^{} > M_N$, 
 secondary vertices can be tagged at the LHC, 
 since right-handed neutrinos can be long-lived.  
In our setup, light neutrinos are considered as Majorana particles, 
 so that lepton number violating processes could also be 
 detectable at ILC.
We will show it especially by analyzing a process of 
 $e^-e^-\to H^-H^-$. 
Anyhow, we stress again that characteristic signals 
 of the $\nu$THDM could be detected 
 at LHC and ILC experiments, and 
 investigating them is our purpose of this paper. 

\section{Setup}

At first, 
 in this section,
 let us show a structure of the $\nu$THDM with $N_R^{}$ and an example 
 of its UV theory. 

\subsection{$\nu$THDM with right-handed Majorana masses}

The 
 $\nu$THDM is 
 originally suggested in Ref.~\cite{Ma}. 
In a framework of this model, 
 Majorana neutrinos were studied in 
 Refs.\cite{Ma, {Ma:2006km}}
 and 
 Dirac neutrinos in 
 Ref.\cite{{Nandi}, {Davidson:2009ha}}. 
In the model, one additional Higgs doublet $\Phi_{\nu}$,
 which gives only neutrino Dirac masses, 
 is introduced 
 besides the SM Higgs doublet $\Phi$ as
\begin{eqnarray}
\Phi_{\nu}=
\left(
\begin{array}{c}
\phi^{+}_{\nu}\\
\phi^{0}_{\nu}
\end{array}
\right),\;\;
\Phi =
\left(
\begin{array}{c}
\phi^{+}\\
\phi^{0}
\end{array}
\right). 
\end{eqnarray}
We introduce a discrete $Z_2$-parity
 which distinguishes $\Phi_\nu$ from $\Phi$,
 and 
 the $Z_2$-charges (and also lepton number) are assigned 
 as the following table. 
\begin{table}[h]
\centering
\begin{center}
\begin{tabular}{|l|c|c|} \hline
fields  &  $Z_{2}$-parity & lepton number \\ \hline\hline
SM Higgs doublet, $\Phi$  &  $+$ &  0 \\ \hline
new Higgs doublet, $\Phi_{\nu}$ 
 &  $-$ & 0 \\ \hline
right-handed neutrinos, $N_R$  &  $-$ & $1$ \\ \hline
others  &  $+$ & $\pm 1$: leptons, $0$: quarks \\ \hline
\end{tabular}
\end{center}
\end{table}

\noindent
Under the discrete symmetry, 
 Yukawa interactions and Majorana mass terms of 
 right-handed neutrinos are given by 
\begin{eqnarray}
-{\mathcal L}_{Yukawa}=\overline{Q}Y_u U_{R}\tilde{\Phi}
 +\overline{Q}Y_d D_{R} \Phi
 +\overline{L}Y_e E_{R} \Phi
 +\overline{L}Y_\nu N_R \tilde{\Phi}_\nu 
+ \frac12\overline{{N_R}^c} M_N N_R + \text{H.c.}, 
\label{Yukawa:nuTHDM}
\end{eqnarray}
where 
 $\tilde{\Phi}=i\sigma_{2}\Phi^{\ast}$, and   
 we omit a generation index. 
We here take the real diagonal bases for charged lepton and right-handed neutrino 
mass matrices without loss of generality. 
An extra scalar doublet $\Phi_\nu$ only couples 
 with $N_R$ by the $Z_2$-parity, so that FCNCs 
 are forbidden at tree level. 
Quark and charged lepton sectors are the same as Type-I THDM~\cite{2hdm}, 
 but notice that 
 this $\nu$THDM is quite different from 
 conventional Type-I, II, X, Y THDMs~\cite{typeX} due to the tiny VEV.

Now let us concentrate on the neutrino sector. 
From Eq.~\eqref{Yukawa:nuTHDM}, the general mass matrix 
 for Majorana neutrino is given by 
\begin{align}
-\frac12
\begin{pmatrix}
\overline{{\nu'_L}^c} & \overline{N'_R}
\end{pmatrix}
\begin{pmatrix} {\bf 0} & \frac{v_\nu}{\sqrt2} Y_\nu^* \\
\frac{v_\nu}{\sqrt2} Y_\nu^\dag & M_N \end{pmatrix}
\begin{pmatrix} \nu'_L\\ {N'_R}^c \end{pmatrix}
+\text{H.c.}, 
\end{align}
where we denote
 $\langle \phi_\nu^0 \rangle = v_\nu/\sqrt{2}$.  
If $\mathcal{O} (Y_\nu v_\nu/M_N) \ll 1$, the matrix can be approximately 
block diagonalized as 
\begin{align}
\begin{pmatrix} \nu'_L\\ {N'_R}^c \end{pmatrix}
= 
\begin{pmatrix} {\bf 1} & + \frac{v_\nu}{\sqrt2} Y_\nu M_N^{-1} \\
- \frac{v_\nu}{\sqrt2} M_N^{-1} Y_\nu^\dag  & {\bf 1} \end{pmatrix}
\begin{pmatrix} U_\text{MNS} \nu_L^{}\\ {N_R}^c \end{pmatrix}. 
\end{align}
where $U_\text{MNS}^{}$ is the neutrino mixing matrix.
Mass eigenstates for light neutrinos are determined 
 by diagonalization of $3\times 3$ block as 
\begin{align}
m_\nu 
&= -\frac{v_\nu^2}2Y_\nu M_N^{-1}Y_\nu^T
= U_\text{MNS} m_\nu^\text{diag} U^T_\text{MNS},
\end{align}
where $m_\nu^\text{diag}$ is real diagonal. 
In order to realize tiny neutrino masses, parameters can be chosen as 
i) small $Y_\nu$, ii) tiny $v_\nu$, or iii) huge $M_N^{}$. 
Here the $\nu$THDM focuses
 on the second possibility which may give interesting Higgs 
and $N_R$ phenomenology due to the relatively large neutrino Yukawa coupling 
and the lighter right-handed neutrinos 
($M_N$ is supposed to be below the TeV scale).
Actually, radiative effects also induce Majorana neutrino masses,
 which are estimated later.

The Higgs potential of the $\nu$THDM is given by 
\begin{align}
V^\text{THDM} 
&
= m_\Phi^2 \Phi^\dag \Phi + m_{\Phi_\nu}^2 \Phi_\nu^\dag \Phi_\nu
-m_3^2\left(\Phi^\dag \Phi_\nu+\Phi_\nu^\dag \Phi\right)
+\frac{\lambda_1}2(\Phi^\dag \Phi)^2
+\frac{\lambda_2}2(\Phi_\nu^\dag \Phi_\nu)^2\nonumber \\
&\qquad+\lambda_3(\Phi^\dag \Phi)(\Phi_\nu^\dag \Phi_\nu)
+\lambda_4(\Phi^\dag \Phi_\nu)(\Phi_\nu^\dag \Phi)
+\frac{\lambda_5}2\left[(\Phi^\dag \Phi_\nu)^2
+(\Phi_\nu^\dag \Phi)^2\right]. 
\label{Eq:HiggsPot}
\end{align}
The $Z_2$-symmetry is softly broken by $m_3^2$. 
Taking a parameter set, 
\begin{equation}
m_\Phi^2 < 0, ~~~ m_{\Phi_\nu}^2 > 0, ~~~ |m_{3}^2| \ll m_{\Phi_\nu}^2,
\end{equation}
and denoting $\langle \phi^0 \rangle = v/\sqrt{2}$, 
 we can obtain the VEV hierarchy of Higgs doublets,
\begin{equation}
v^2 \simeq \frac{-2m_\Phi^2}{\lambda_1},
 ~~~ v_{\nu} \simeq \frac{-m_{3}^2 v}{ m_{\Phi_\nu}^2 + (\lambda_3 + \lambda_4 + \lambda_5 ) v^2/2} .
\end{equation}
When we take values of parameters as 
 $m_\Phi \sim 100$ GeV, 
 $m_{\Phi_\nu} \sim 1$ TeV, 
 and $|m_{3}^2| \sim 10$ GeV$^2$, 
 we can obtain $v_\nu \sim 1$ MeV. 
The smallness of $|m_{3}^2|$ is guaranteed by the 
 ``softly-broken'' $Z_2$-symmetry. 
This VEV hierarchy is preserved after estimating a 
 one-loop effective potential as long as 
 $m_{\Phi_\nu}$ is larger than weak scale, 
 since modifications of stationary conditions are
 negligible.

For a large $\tan \beta=v/v_{\nu} (\gg 1)$ limit we are interested in,
 five physical Higgs states and those
 masses are given by
\begin{eqnarray}
 h \simeq {\rm Re} [\phi^0] ,          && ~~~ m^2_{h} \simeq 2\lambda_1 v^2  , \\ 
 H \simeq {\rm Re} [\phi_\nu^0] ,      && ~~~ m^2_{H} \simeq
  m_{\Phi_\nu}^2 + {1\over2}(\lambda_3 + \lambda_4+\lambda_5) v^2 , \\
 A \simeq  {\rm Im} [\phi_{\nu}^0] ,   && ~~~ m^2_A \simeq m_{\Phi_\nu}^2
  +{1\over 2}
  (\lambda_3 + \lambda_4 - \lambda_5) v^2 ,       \\
 H^\pm \simeq \ [\phi_\nu^\pm] ,       && ~~~ m^2_{H^\pm} \simeq
  m_{\Phi_\nu}^2
 + \frac{1}{2}\lambda_3 v^2 ,
\end{eqnarray}
respectively, 
 where we omit negligible corrections of
 ${\cal O}(v_{\nu}^2)$ and ${\cal O}(m_3^2)$. 
Notice that 
 $\tan\beta$ is extremely large, so that 
 the SM-like Higgs boson $h$ is almost originated from $\Phi$,
 while other physical Higgs particles, $H^\pm, H, A$, are almost
 originated from $\Phi_\nu$. 
Since $\Phi_\nu$ has Yukawa couplings only with right-handed neutrinos and 
 lepton doublets, 
 remarkable phenomenology can be expected which 
 are not observed in other THDMs. 
As early works, 
 lepton flavor violation (LFV) processes 
 and oblique corrections in $\nu$THDM are 
 estimated in Ref.~\cite{Ma}, and  
 charged Higgs boson phenomenology in collider experiments
 are discussed 
 in Refs.~\cite{Davidson:2009ha, Logan:2010ag}~\footnote{
The model\cite{Davidson:2009ha, Logan:2010ag}
 deals with Dirac neutrino version
 in $\nu$THDM, and  
 phenomenology of charged Higgs
 has a similar region in part.}.

\begin{figure}[tb]
\centering
\includegraphics[height=4.5cm]{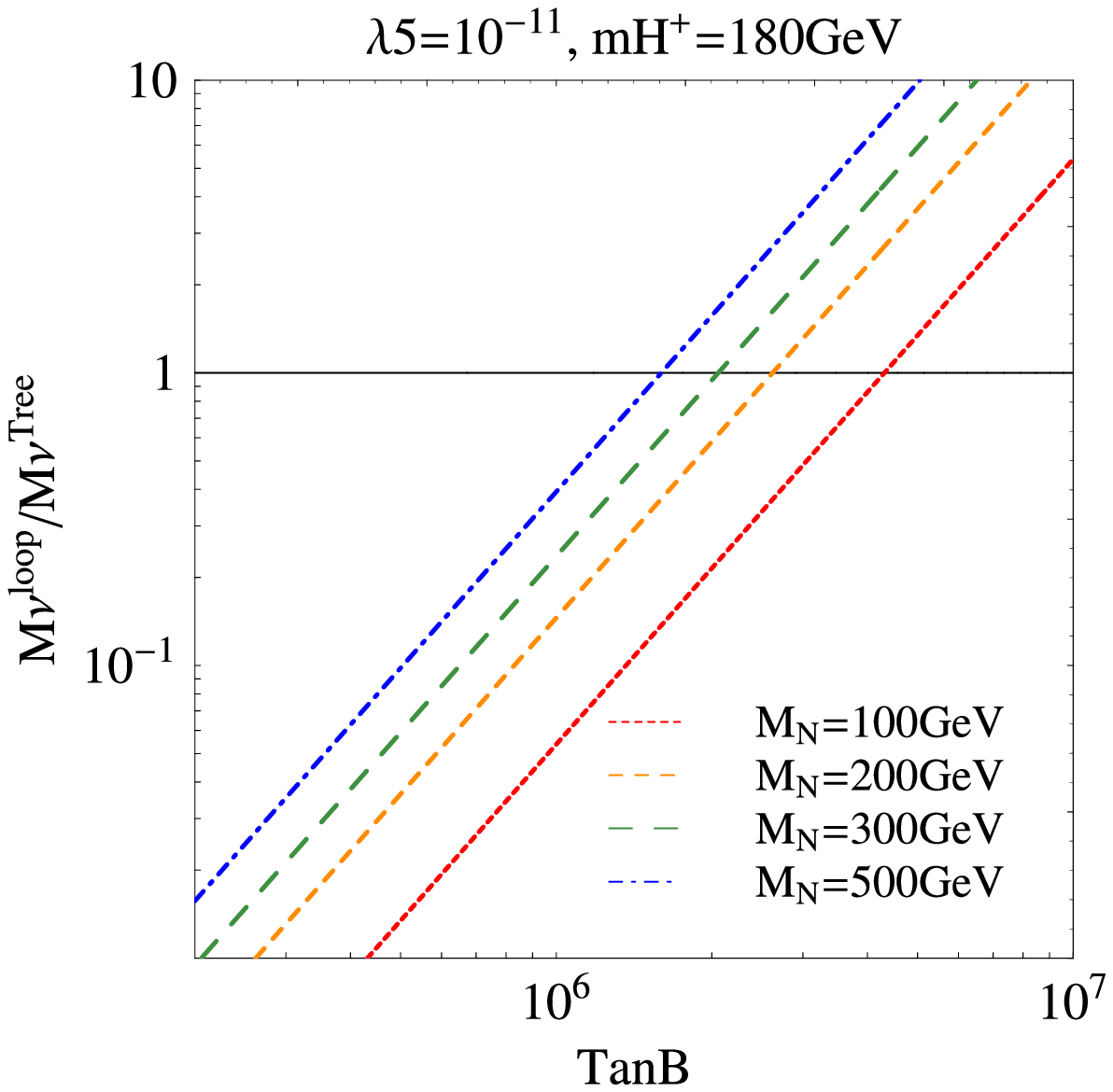}
\includegraphics[height=4.5cm]{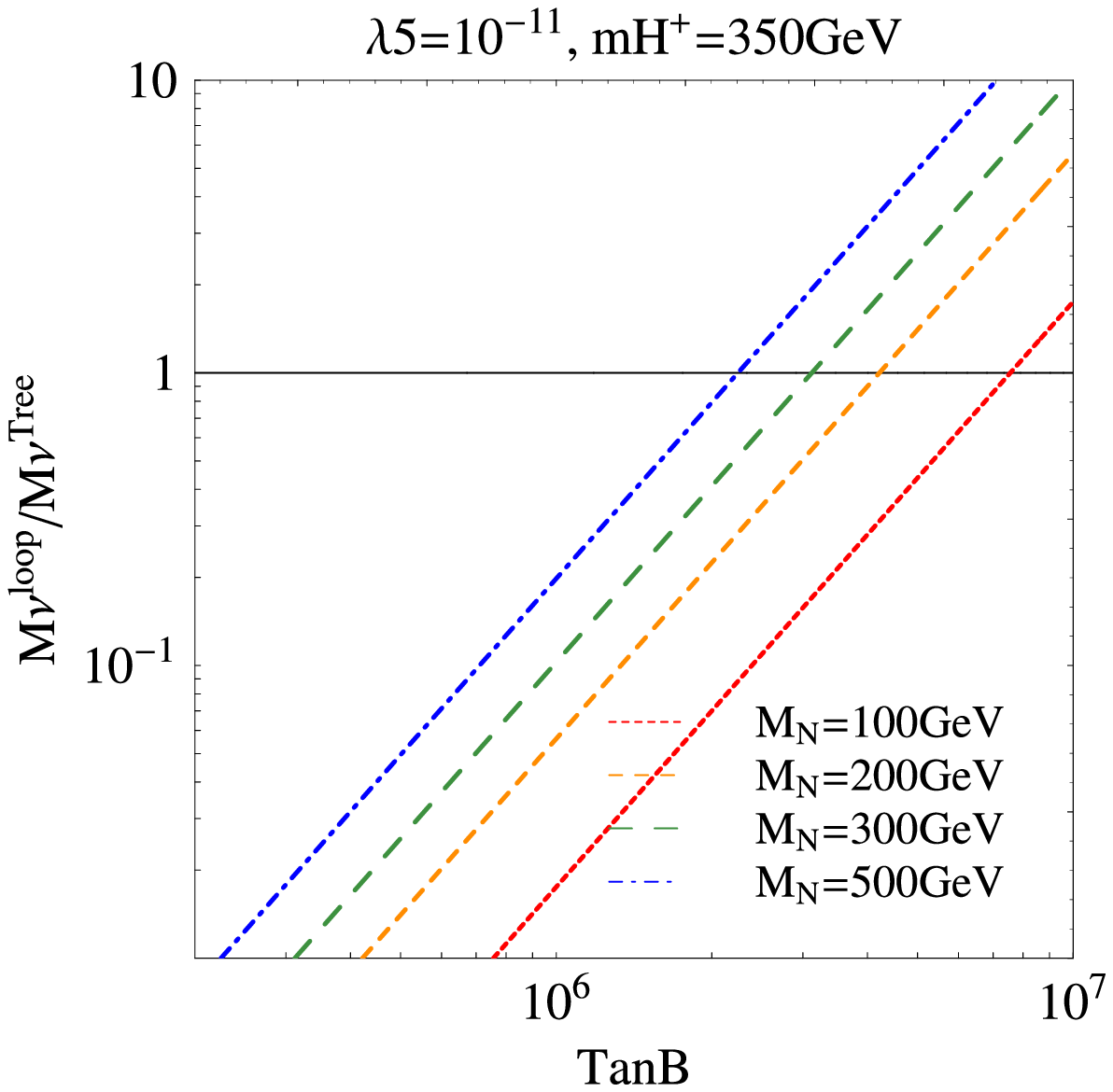}
\caption{The ratio of the tree level and the loop level contributions 
to the neutrino mass.
}
\label{FIG:ratio}
\end{figure}

Now we are in a position to estimate 
 the neutrino mass from one-loop radiative
 corrections. 
There are one-loop diagrams which induce
 Majorana neutrino masses\cite{Ma:2006km}. 
The effects exist when $\lambda_5 \neq 0$, and 
 the dimension five operator is induced. 
It is estimated as 
\begin{eqnarray}
(m_\nu^\text{loop})_{ij} &&= -\sum_k {(Y_\nu)_{ik} (Y_\nu)_{jk} M_{Rk} \over 16
 \pi^{2}} 
\left[ 
{m_H^{2} \over m_H^{2}-M_{Rk}^{2}} \ln {m_H^{2} \over M_{Rk}^{2}} - 
{m_A^{2} \over m_A^{2}-M_{Rk}^{2}} \ln {m_A^{2} \over M_{Rk}^{2}}
\right] \nonumber \\
&&\simeq -{\lambda_5 v^{2} \over 8 \pi^{2}} 
\sum_k {(Y_\nu)_{ik} (Y_\nu)_{jk} M_{Rk} \over m_0^{2} - M_{Rk}^{2}} \left[ 
1 - {M_{Rk}^{2} \over m_0^{2}-M_{Rk}^{2}} \ln {m_0^{2} \over M_{Rk}^{2}}
 \right],
\label{216}
\end{eqnarray}
where
 $m_H^{2} - m_A^{2} = \lambda_5 v^{2} \ll m_0^{2} = (m_H^{2} + m_A^{2})/2$. 
If the masses of Higgs bosons (except for $h$) and right-handed 
 neutrinos are ${\mathcal O}(1)$ TeV, light neutrino mass scale of order 
 ${\mathcal O}(0.1)$ eV is induced with $\lambda_5 \sim 10^{-4}$ for 
 $Y_\nu \sim 10^{-3}$.
Whether tree-level effect is larger than loop-level effect 
 or not is determined by the magnitude of $\lambda_5$. 
Roughly speaking, when $\lambda_5 v^2/(4\pi)^2 \gtrsim v_\nu^2$, 
 radiatively induced mass would be dominated. 
If we consider the Type-I seesaw dominance, 
 $v_\nu$ would be a valuable parameter. 
By fixing the light neutrino mass and $M_N$, 
 neutrino data can be fit for wide range of 
 parameter space in $v_\nu$ 
 with a change the magnitudes of $Y_\nu$. 
And, since the magnitude of $v_\nu$ is determined 
 by $m_3^2$, it is the essential parameter in $\nu$THDM. 
We show the ratio of contributions to the light Majorana neutrino mass from the 
tree level and radiative seesaws as a function of $\tan\beta$ in FIG.~\ref{FIG:ratio}. 
Two models, vanishing $v_\nu$\cite{Ma:2006km} and
 non-vanishing $v_\nu$\cite{Ma},
 are connected by the magnitude of $v_\nu$. 
The former model has exact $Z_2$-parity 
 with $m_3^2=0$, where right-handed neutrino 
 becomes dark matter and 
 masses of light active neutrinos are fully 
 induced by the radiative corrections as discussed above.   

\subsection{An example of UV theory of $\nu$THDM}

Next,
 let us show an example of UV theory\cite{HabaHirotsu} of the 
 $\nu$THDM. 
This model is constructed by introducing
 one gauge singlet scalar field $S$, 
 which has a lepton number,
 and $Z_3$-charges are shown 
 as the following table. 
\begin{table}[h]
\centering
\begin{center}
\begin{tabular}{|l|c|c|} \hline
fields  &  $Z_{3}$-charge & lepton number \\ \hline\hline
SM Higgs doublet, $\Phi$  &  1 &  0 \\ \hline
new Higgs doublet, $\Phi_{\nu}$ 
 &  $\omega^{2}$ & 0 \\ \hline
new scalar singlet, $S$  &  $\omega$  & 
 $-2$ \\ \hline
right-handed neutrinos, $N_R$  &  $\omega$ & 1 \\ \hline
others  &  1 & $\pm1$: leptons, $0$: quarks \\ \hline
\end{tabular}
\end{center}
\end{table}
\noindent
Under the discrete symmetry, 
 Yukawa interactions are given by 
\begin{eqnarray}
-{\mathcal L}_{Yukawa}=\overline{Q}Y_u U_{R}\tilde{\Phi}
 +\overline{Q}Y_d D_{R}\Phi
 +\overline{L}Y_e E_{R}\Phi 
 +\overline{L}Y_\nu N_R \tilde{\Phi}_{\nu}
 +\frac12 SY_N\overline{N_R{}^c}N_R +\text{H.c.}\;
\label{22}
\end{eqnarray}
Notice that 
 the effective Majorana masses are induced 
 in the $\nu$THDM as  
\begin{equation}
M_R = {Y_{N}\langle S \rangle}. 
\end{equation}
The Higgs potential can be written as
\begin{eqnarray}
V=&m_\Phi^{2}|\Phi|^{2}+m_{\Phi_{\nu}}^{2}|\Phi_{\nu}|^{2}-m_S^{2}|S|^{2}
  +\frac{\lambda_{1}}{2}|\Phi|^{4}
  +\frac{\lambda_{2}}{2}|\Phi_{\nu}|^{4}
  +\lambda_{3}|\Phi|^{2}|\Phi_{\nu}|^{2} 
  +\lambda_{4}|\Phi^{\dagger}\Phi_{\nu}|^{2} \nonumber \\
  &
   +\lambda_{S}|S|^4+
  \lambda_{\Phi}|S|^{2}|\Phi|^{2}
  +\lambda_{\Phi_{\nu}}|S|^{2}|\Phi_{\nu}|^{2} 
  + \Bigl[ -\lambda S^{3}-\kappa S\Phi^{\dagger}\Phi_{\nu}
 + \text{H.c.} \Bigr].
\label{Potential:HabaHirotsu}
\end{eqnarray}
$Z_3$-symmetry forbids 
 dimension four operators, 
 $(\Phi^\dagger \Phi_{\nu})^{2}$, 
 $\Phi^\dagger \Phi_{\nu}|\Phi|^2$, 
 $\Phi^\dagger \Phi_{\nu}|\Phi_{\nu}|^2$,
 $S^4$, $S^2|S|^2$, $S^2|\Phi|^2$,
 $S^2|\Phi_{\nu}|^2$, 
 and dimension two or three operators, 
 $\Phi^\dagger \Phi_{\nu}$, 
 $S|\Phi|^{2}$, $S|\Phi_\nu|^{2}$. 
Although there might be introduced small soft breaking 
 terms such as $m_3'^2\Phi^\dagger \Phi_{\nu}$ to avoid 
 domain wall problem, 
 we omit them here, for simplicity. 
It has been shown that, with $\kappa \sim 1$ MeV,
 the desirable hierarchy of 
 VEVs 
\begin{eqnarray}
&& v_s \equiv \langle S \rangle \sim  1 \;\hbox{TeV},\;\;\;
  v \sim  100 \;\hbox{GeV}, \;\;\;
  v_{\nu} \sim  1 \;\hbox{MeV}, 
\end{eqnarray}
and neutrino mass  
 $m_\nu \sim \frac{Y_{\nu}^2  v_{\nu}^{2}} {M_R}$
 can be realized~\cite{HabaHirotsu}. 
This is so-called 
 Type-I seesaw mechanism in a TeV scale, when
 coefficients 
 $Y_\nu$ and $Y_N$ are assumed to be 
 of order one. 
The masses of scalar and pseudo-scalar mostly from VEV of $S$ are
 given by
\begin{eqnarray}
 m^{2}_{H_{S}} &=& m_S^{2}+2\lambda_{S} v_s^2 , \;\;\;\;\;\;
 m^{2}_{A_{S}} = 9 \lambda v_s ,
\end{eqnarray} 
where we assume the CP invariant Higgs sector. 
For parameter region with $v_s \gg 1$ TeV, 
 both scalar and pseudo-scalar are heavier than other particles.
After integrating out $S$, thanks to the $Z_3$-symmetry, 
 the model ends up with an effectively $\nu$THDM 
 with approximated $Z_2$-symmetry, $\Phi \to \Phi, \Phi_\nu \to -\Phi_\nu$. 
Comparing to the $\nu$THDM, 
 the value of $m_3^2$, which is a soft $Z_2$-symmetry breaking
 term, is expected to be 
 $\kappa v_s$. 
$\lambda_{5}$ is induced by integrating out $S$, which is 
 estimated as ${\mathcal O}(\kappa^2/m_S^2)\sim 10^{-12}$.

As for the neutrino mass induced from one-loop diagram, 
 UV theory induces small $\lambda_5\sim 10^{-12}$  
 as shown above, 
 then the radiatively induced neutrino mass from 
 one-loop diagram
 is estimated as $\lambda_5 v^2/(4\pi)^2M \sim 10^{-4}$
 eV. 
This can be negligible comparing to light neutrino mass
 which is induced from 
 tree level Type-I seesaw mechanism. 
The tree level neutrino mass is 
\begin{equation}
m_\nu^{tree}
 \sim {y_\nu^2 v_\nu^2 \over M}\sim 
{y_\nu^2 \kappa^2v^2 \over v_s^2M},
\end{equation}
where we input 
 $v_\nu \sim {\kappa v \over v_s}$.  
On the other hand, one-loop induced neutrino mass is 
 estimated as 
\begin{equation}
m_\nu^{loop} \sim {\lambda_5 y_\nu^2 \over (4\pi)^2}{v^2 \over M}
\sim {y_\nu^2 \over (4\pi)^2}{\kappa^2 v^2 \over M^2 M}.
\end{equation}
Putting $M\sim v_s$, 
\begin{equation}
{m_\nu^{loop} \over m_\nu^{tree}}\sim 
{1 \over (4\pi)^2},
\end{equation} 
which shows loop induced neutrino mass
 is always smaller than tree level mass
 if UV theory is the model of Ref.~\cite{HabaHirotsu}.

\section{Phenomenology in $\nu$THDM with $N_R$}

In this section, we study phenomenology 
 in the $\nu$THDM with $N_R$, which makes a low energy 
 seesaw mechanism work.  
At first, we will briefly show experimental 
 constraints of the $\nu$THDM. 
And next, we will show observable collider phenomenology
 in LHC and ILC experiments. 

\subsection{Experimental constraints}

Let us summarize experimental constraints on the model, and 
the first is LFV.
The muon LFV is precisely measured, 
 and MEGA experiment gives a strong upper limit as 
 ${\mathcal B}_{\mu\to e\gamma} < 1.2 \times 10^{-11}$\cite{Brooks:1999pu}. 
The branching fraction of radiative LFV decay 
 due to neutrinophilic charged Higgs boson is calculated as\cite{Ma}
\begin{align}
{\mathcal B}_{\mu\to e\gamma} = \frac{3\alpha_\text{EM}^{}}{2\pi}
\Bigl|\frac{v^2}{2m_{H^\pm}^2} (Y_\nu)_{e N} (Y_\nu)_{\mu N}^*
F_2(\tfrac{M_N^2}{m_{H^\pm}^2}) \Bigr|^2,
\end{align}
where 
\begin{align}
F_2(t) = \frac1{12(1-t)^4}(1-6t+3t^2+2t^3-6t^2\ln t).
\end{align}
Assuming masses of charged Higgs boson and right-handed 
 neutrino to be TeV scale, $Y_\nu \lesssim 0.1$ 
 should be satisfied. 
When the neutrino mass dominantly induced by Type-I seesaw mechanism, 
 the neutrino Yukawa coupling matrix can be expressed as\cite{casasibara}
\begin{align}
Y_\nu \simeq i\, \frac{\sqrt2}{v_\nu} U_\text{MNS} 
{\widehat m_\nu}^{1/2} R {\widehat M_N}^{1/2}
=i\, \frac{\sqrt2}{v} U_\text{MNS} 
{\widehat m_\nu}^{1/2} R {\widehat M_N}^{1/2} \tan\beta,
\label{casasibara}
\end{align}
where $R$ is a complex orthogonal matrix. Small $v_\nu$, 
 equivalently large $\tan\beta$, is constrained by 
 $\mu \to e\gamma$ results. 
However, we note that small $v_\nu$ is not excluded,  
 because the one-loop contribution to neutrino mass is 
 dominated in this case (see FIG. 1), and hence
 Eq.\eqref{casasibara} is no longer valid.

Next is about $\rho$-parameter, and 
 it often gives severe constraints for models of beyond the SM\cite{2hdm}. 
Since $\rho$-parameter characterizes breaking of custodial 
 $SU(2)$ symmetry, we here simply requires mass degeneracy of 
 extra Higgs bosons, i.e., $m_A^{} \simeq m_{H^\pm}^{}$. 
This assumption is automatically satisfied 
 when $m_{\Phi_\nu}^2 \gtrsim v^2$. 

The last constraint is neutrinoless double beta decay. 
As in the SM, masses of light Majorana neutrinos are 
 the source of neutrinoless double beta decay\cite{doublebeta}. 
In the $\nu$THDM, charged Higgs bosons are {\it quarkophobic}, 
so that it cannot affect to neutrinoless double beta decay. 

\subsection{LHC phenomenology}

Now let us investigate LHC physics of the $\nu$THDM. 
For this purpose, it is good for us to focus on
 decays of charged Higgs bosons and right-handed
 neutrinos. 
It is because 
 we can expect the charged Higgs boson production at the LHC, 
 and its decay processes can be observable\cite{DYCh}.
The charged Higgs boson in the $\nu$THDM is almost composed 
 by $\Phi_\nu$ whose couplings to quarks are negligibly small. 
Therefore, different phenomenology from other THDMs 
 are expected to be observed in accelerator experiments 
 especially in the processes related with 
 charged Higgs particle and 
 right-handed neutrinos which have significant 
 Yukawa interactions.   
Here let us summarize decays of 
 charged Higgs bosons and also right-handed 
 neutrinos, depending on mass spectrum. 

\subsubsection{$m_{H^\pm} < M_N$} 
For $m_{H^\pm}^{} < M_N$, $N_R$ decays into leptons and on-shell 
 scalar bosons via the neutrino Yukawa interaction as 
\begin{align}
\Gamma(N_{Ra}\to \ell^+ H^-)
&= \frac{|(Y_\nu)_{\ell a}|^2 m_{H^\pm}^2}{16\pi M_a} 
\left( 1-\frac{m_{H^\pm}^2}{M_a^2} \right)^2, \\
\Gamma(N_{Ra}\to \nu_i \phi^0_\nu) 
&= \frac{|(U_\text{MNS}^\dag Y_\nu)_{i a}|^2 m_{\phi_\nu^0}^2}{16\pi M_a} 
\left( 1-\frac{m_{\phi_\nu^0}^2}{M_a^2} \right)^2,
\end{align}
where $\phi_\nu^0=H,A$, and lepton masses are neglected.
However, it would be difficult to produce $N_R$ at the LHC
 even when their masses are below $1$ TeV, because 
 right-handed neutrinos are originally gauge singlet, 
 and their weak interactions is suppressed by $Y_\nu v_\nu/M_N$. 

On the other hand, charged Higgs bosons can easily be produced 
 at the LHC via the scalar pair creation process 
 $q \bar q \to \gamma^*(Z^*) \to H^+ H^-$\cite{DYCh}. 
For large $\tan\beta=v/v_\nu$, the main decay modes of 
 charged Higgs boson would be light neutrinos 
 and charged leptons via the Yukawa interaction. 
Because of small mixing between light- and heavy- neutrinos, 
the decay width is strongly suppressed as
\begin{align}
\Gamma(H^+\to \ell^+ {\nu_L^{}}_i)
&= \frac{|(Y_\nu M_N^{-1} Y_\nu^T U_\text{MNS}^*)_{\ell i}|^2 m_{H^\pm}
}{32\sqrt2 \pi G_F \tan^2\beta}.
\end{align}
If $\tan\beta$ is not so large, small mixing in scalar sector 
becomes important. In this case, charged Higgs boson can also 
decay into quarks. 

\begin{figure}[tb]
\centering
\includegraphics[height=4cm]{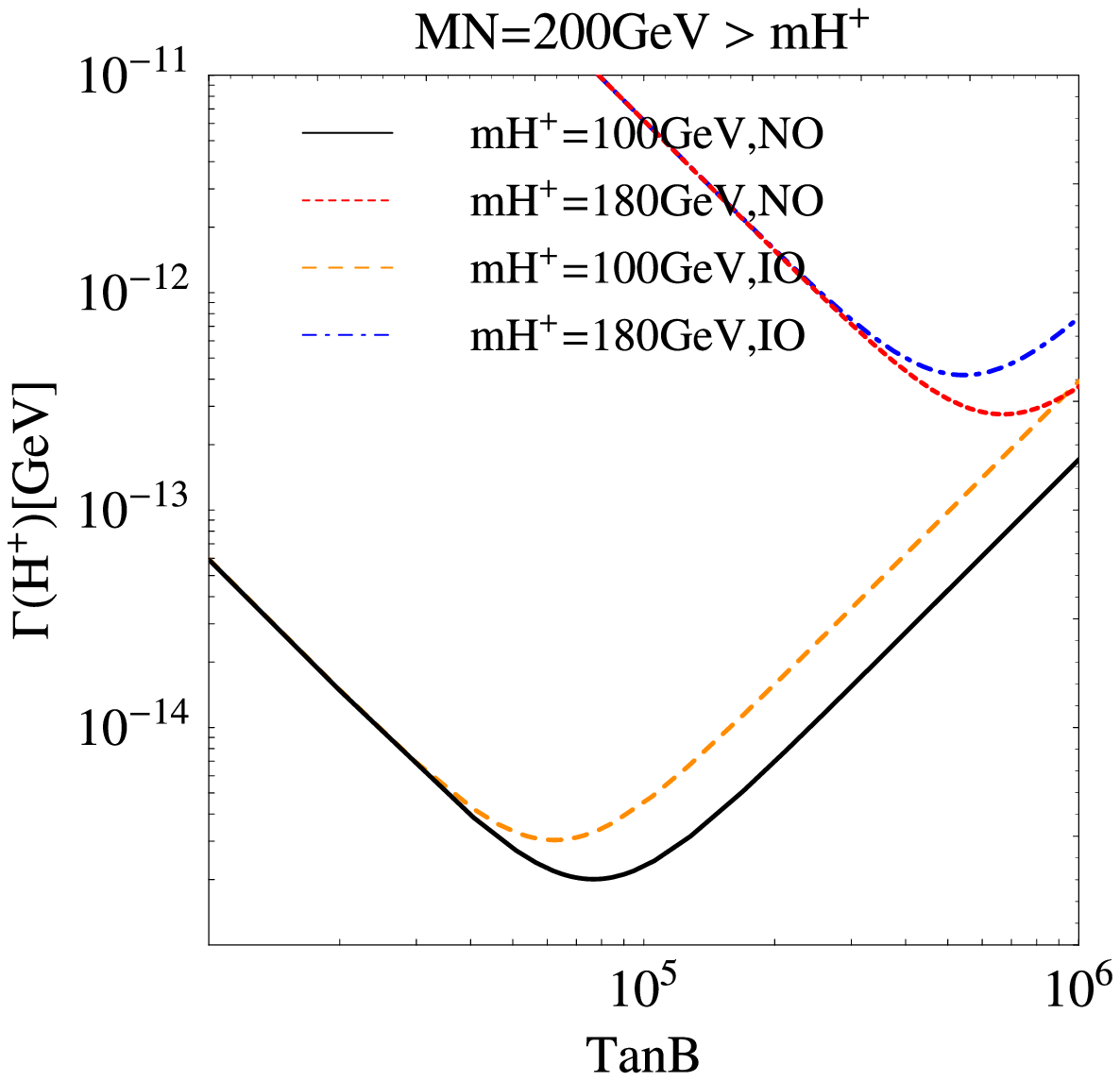}
\includegraphics[height=4cm]{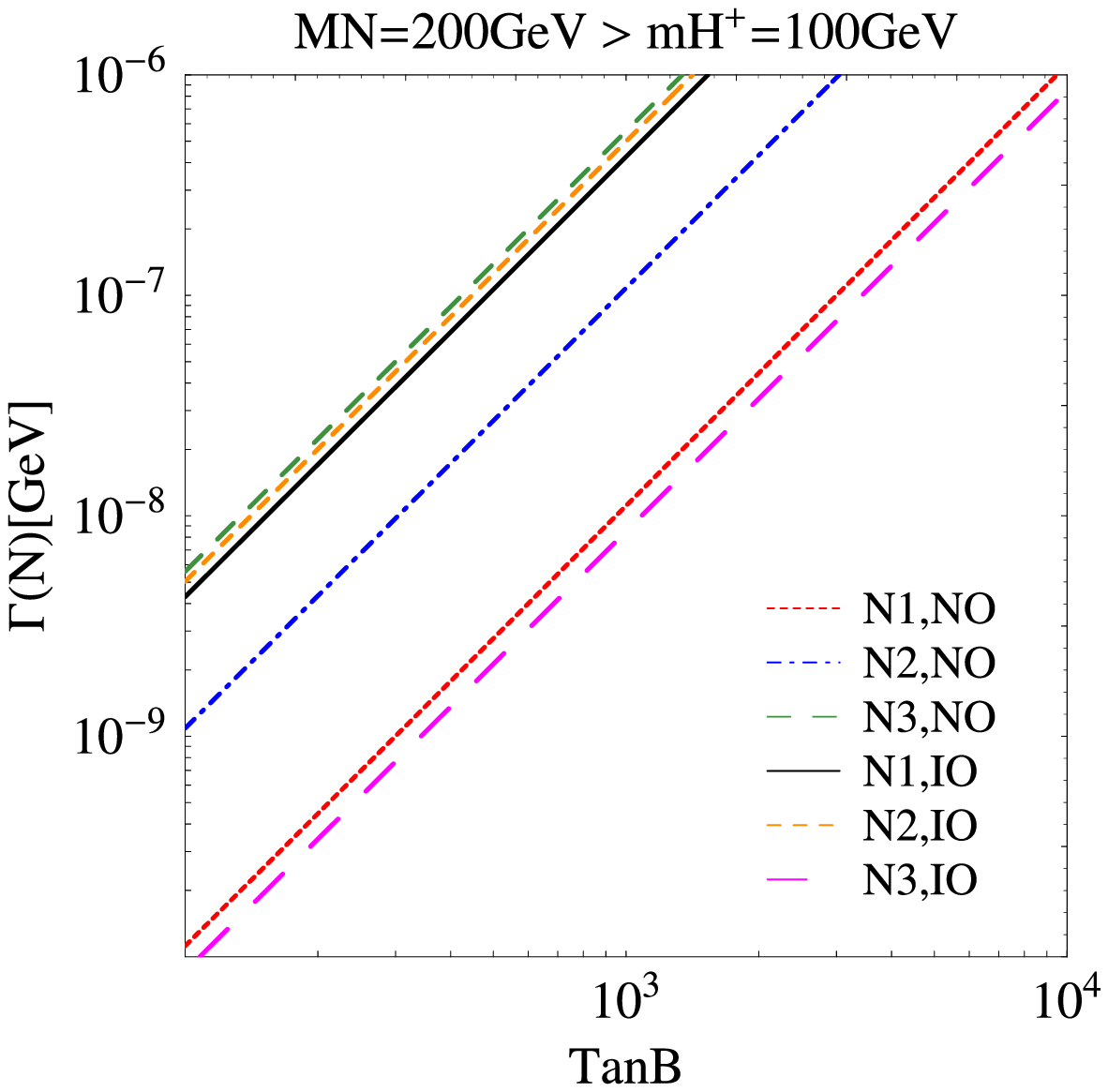}
\includegraphics[height=4cm]{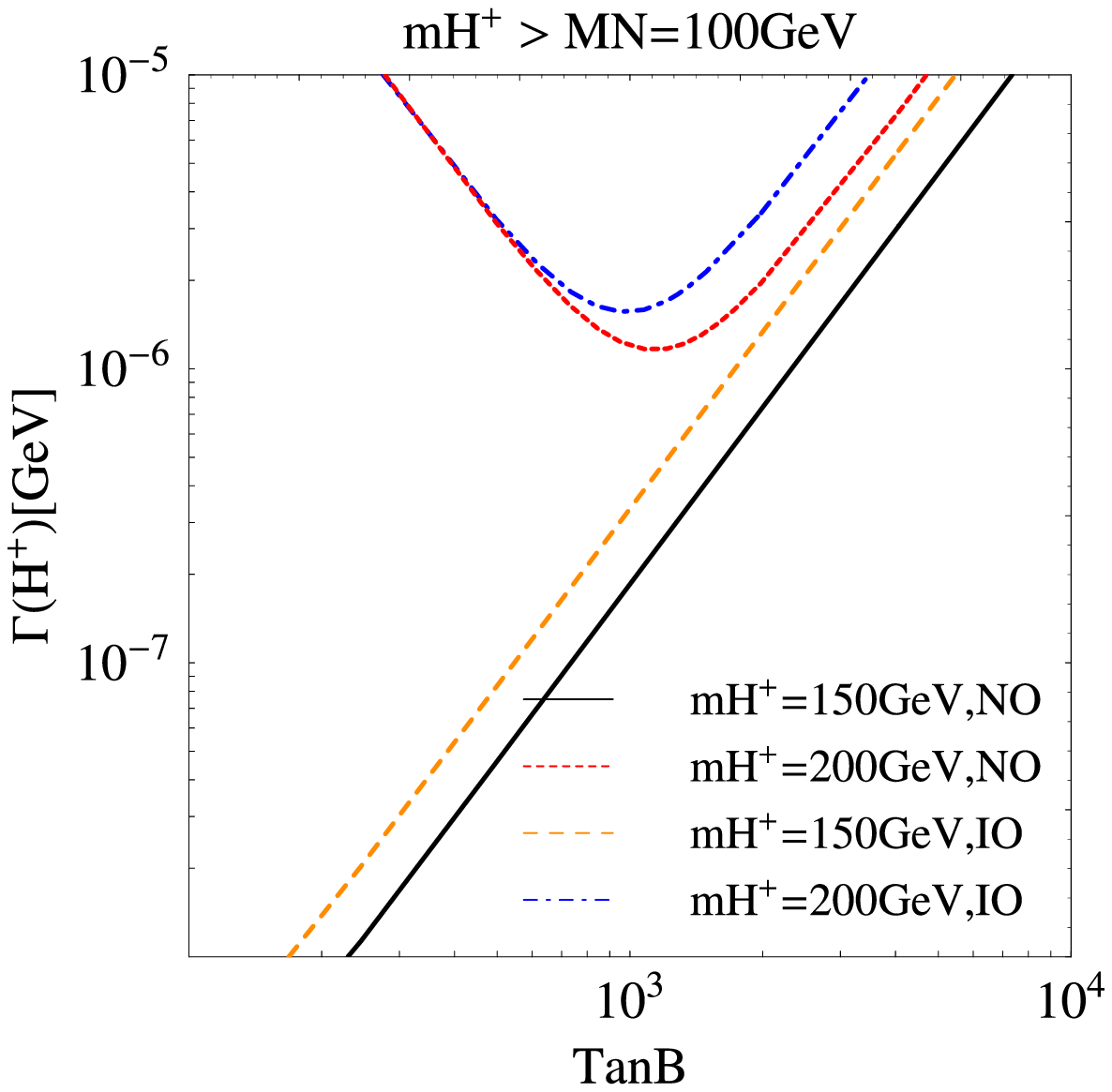}
\includegraphics[height=4cm]{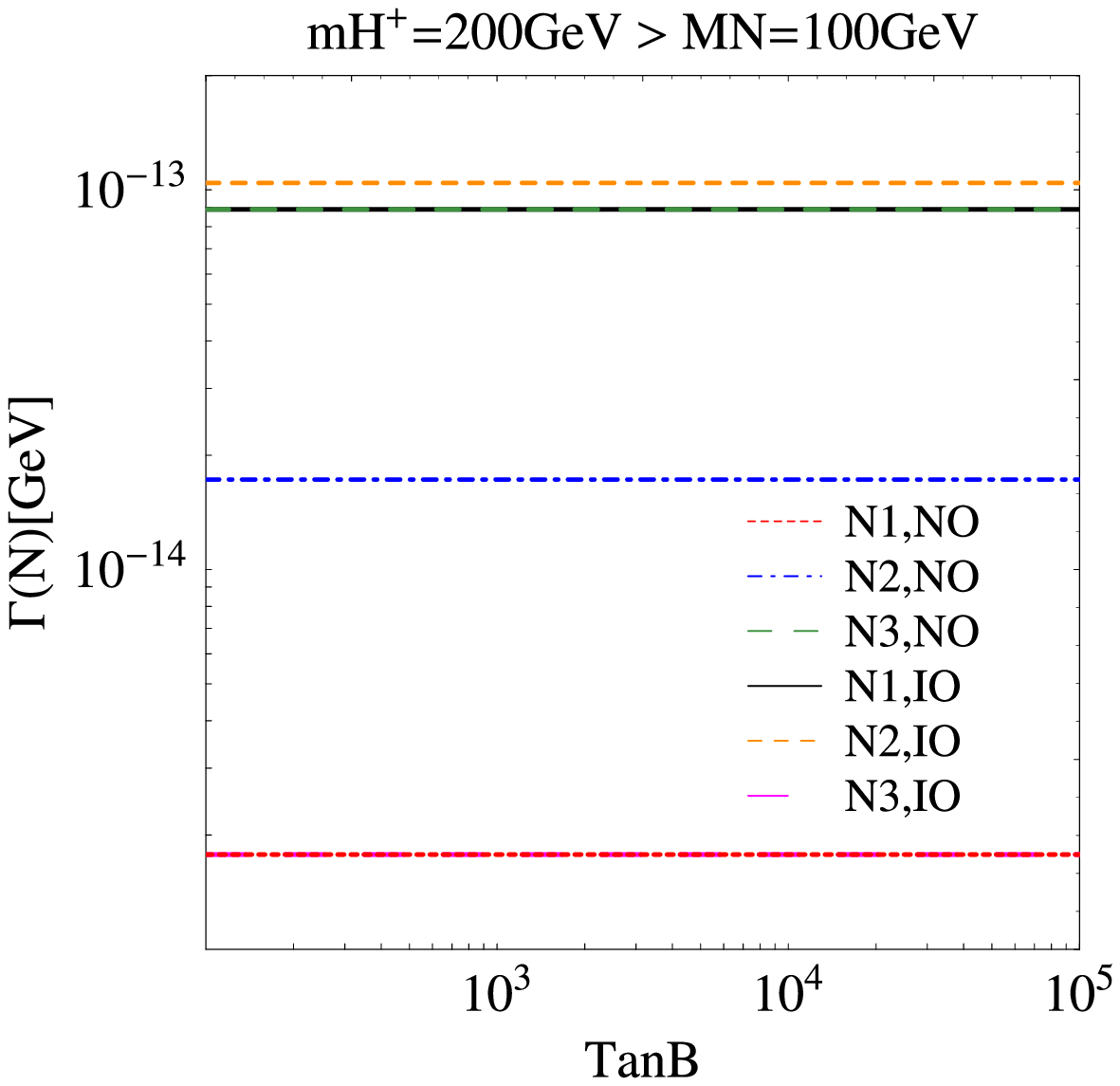}
\caption{Decay widths for the charged Higgs boson and right-handed neutrinos are shown, 
where $R={\bf 1}$ and $M_N \propto {\bf 1}$ are taken. 
}
\label{FIG:width}
\end{figure}
In the first panel of FIG.~\ref{FIG:width}, 
 we show the total decay width of charged Higgs boson. 
We here take the charged Higgs boson mass to be $100$ GeV and $180$ GeV,
 and $R={\bf 1}$ and $M_N \propto {\bf 1}$ are assumed. 
Mass spectrum of light neutrinos is considered for both normal mass 
 ordering and inverted mass ordering case. 
It must be stressed that 
 the decay length of charged Higgs bosons is found to be longer, 
 so that long lived charged Higgs boson can be 
 found at the LHC. 
If the charged Higgs boson is heavier than the top-quark, 
 charged Higgs boson can also decay into $tb$ pairs. 
In this case, the lifetime of charged Higgs bosons is shorter by 
factor of $m_c^2/m_t^2$. 
In the second panel, 
 we also show the total decay width of right-handed Majorana neutrinos. 
Heavy neutrinos instantaneously decay into leptons and scalars. 
Outgoing charged leptons can be tagged if right-handed neutrinos are produced. 

\begin{figure}[tb]
\centering
\includegraphics[height=4cm]{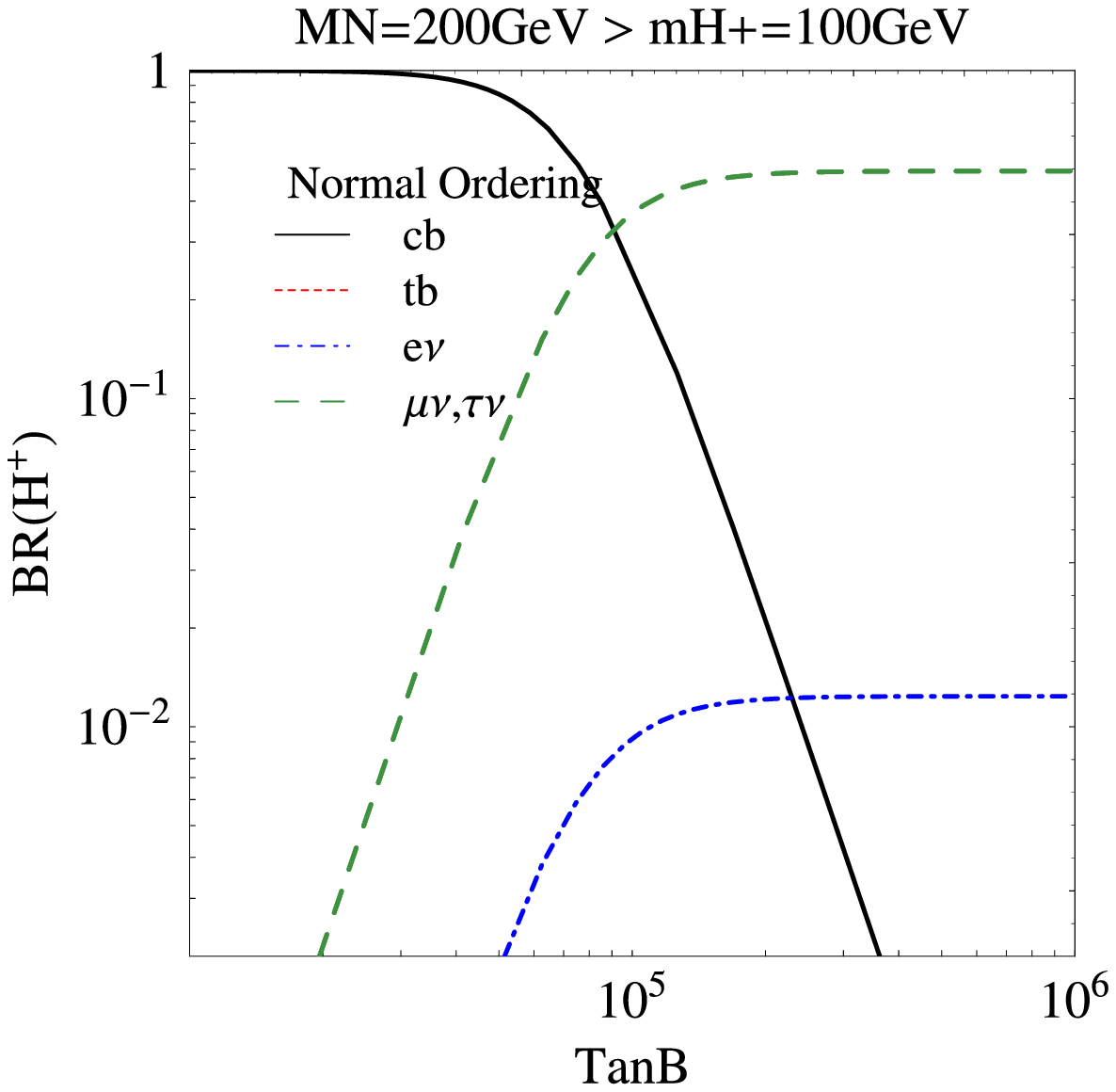}
\includegraphics[height=4cm]{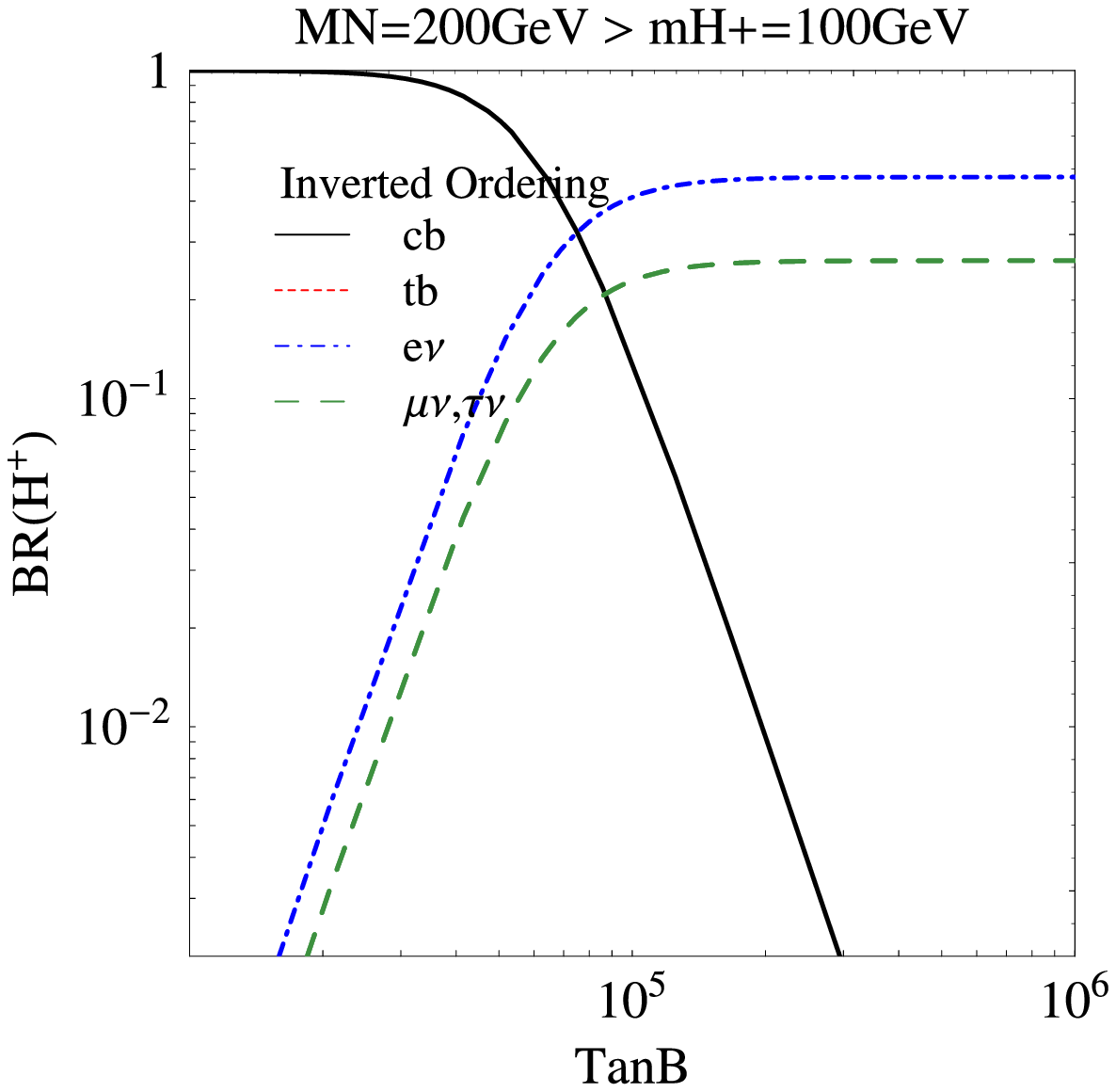}
\includegraphics[height=4cm]{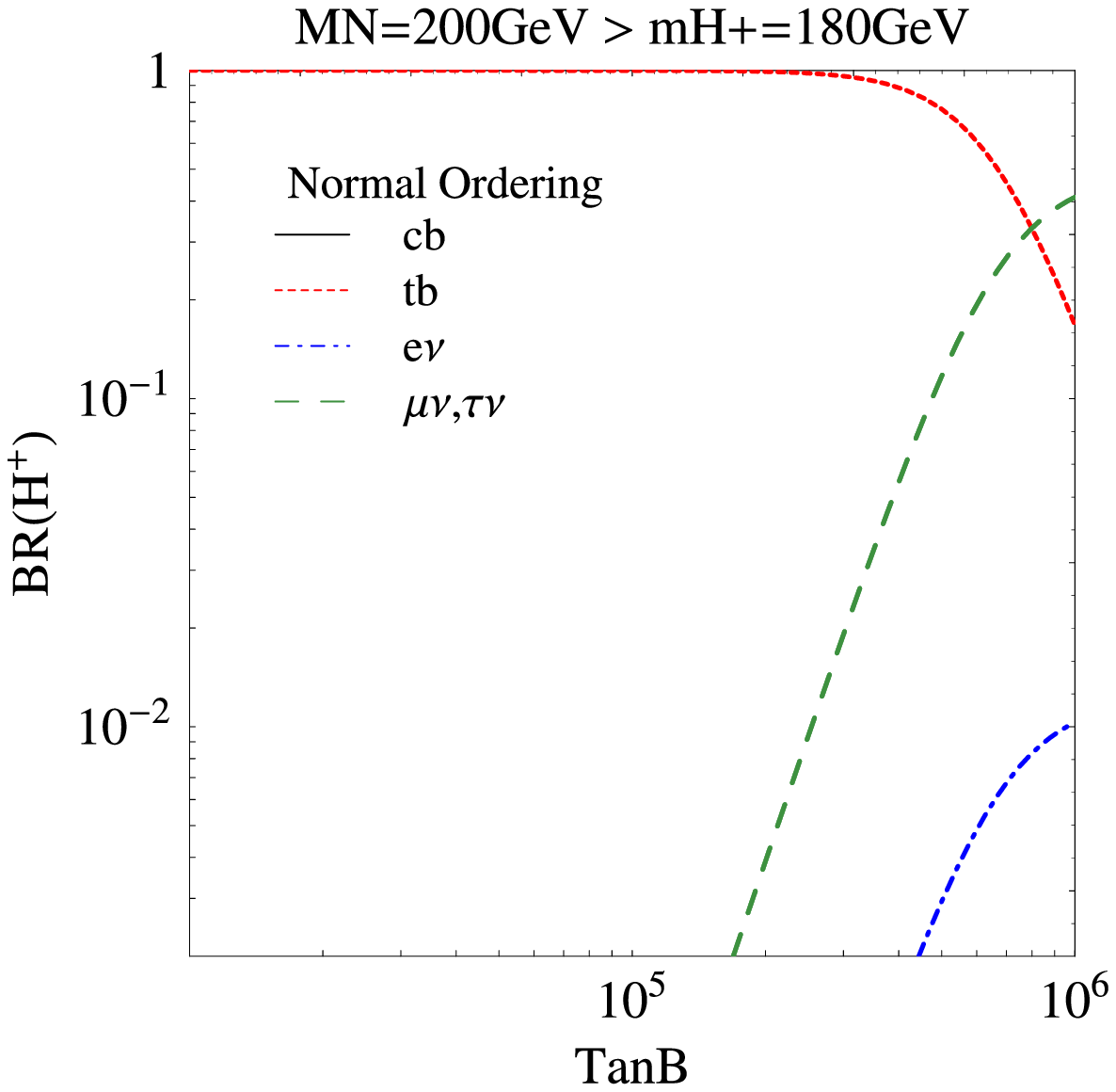}
\includegraphics[height=4cm]{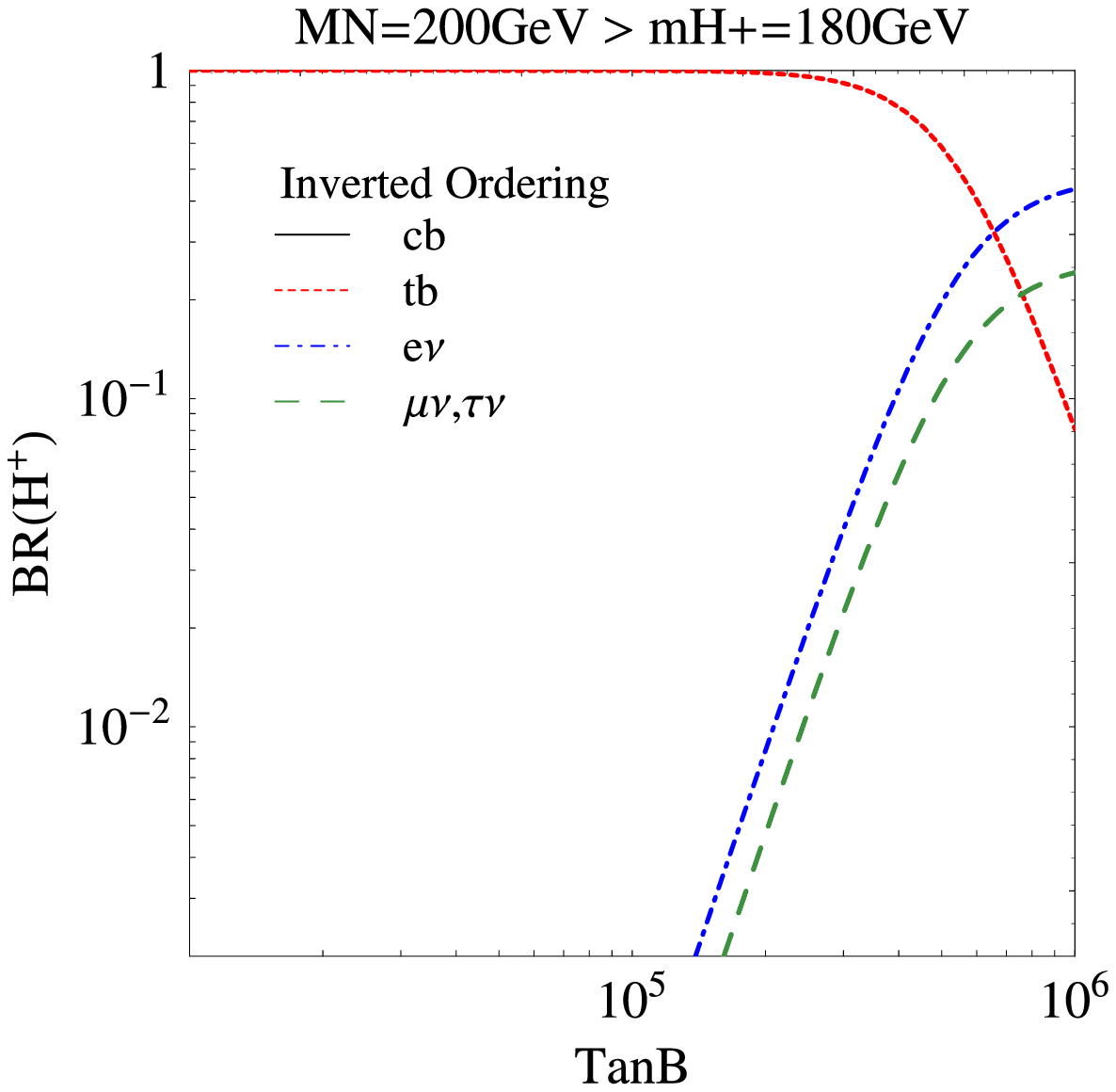}
\caption{Decay branching ratios of the charged Higgs boson are given for $m_{H^\pm}^{} < M_N$, 
where $R={\bf 1}$ and $M_N \propto {\bf 1}$ are taken.
}
\label{FIG:BrChB}
\end{figure}
In FIG.~\ref{FIG:BrChB}, 
 decay branching ratios of charged Higgs boson are shown,  
 where $m_{H^\pm}^{}=100$ GeV (first and second) and $180$ GeV (third and fourth) are taken, 
 and neutrino mass spectrum are assumed to be
 normal mass ordering and inverted mass ordering. 
For lower $\tan\beta$ region, charged Higgs bosons mainly decay 
 into quarks, while for larger $\tan\beta$ region they do into leptons. 
In large $\tan\beta$ region, main leptonic decay channels are 
 different in the normal ordering case and in the inverted ordering 
case\footnote{This prediction would be strongly dependent on the assumption 
of $R$ and right-handed neutrino mass spectrum. However, we note that 
something more information about neutrino Yukawa coupling matrix can 
be obtained by analyzing the charged Higgs decays at colliders.}.

\subsubsection{$m_{H^\pm} > M_N$} 

For $m_{H^\pm}^{} > M_N$, 
 charged Higgs bosons cannot be long-lived.
The
 charged Higgs bosons $(H^+)$ rapidly decay into 
 charged leptons $(\ell^+)$ and on-shell right-handed neutrinos $N_R$ 
 via the neutrino Yukawa interaction as 
\begin{align}
\Gamma(H^+\to \ell^+N_{Ra}) 
&= \frac{|(Y_\nu)_{\ell a}|^2m_{H^\pm}^{}}{16\pi} 
\left( 1-\frac{M_a^2}{m_{H^\pm}^2} \right)^2. 
\end{align}
As we mentioned before, charged Higgs boson can easily be produced 
 at the LHC, and then right-handed neutrinos can be created from their decays. 
Decays of $H$ and $A$ are also used to generate right-handed neutrinos. 
Subsequently, $N_R$ decays into leptons and gauge bosons as 
\begin{align}
\Gamma(N_{Ra}\to \ell^+W^-)
&= \frac{\left|(Y_\nu)_{\ell a}\right|^2 M_a}{32\pi \tan^2\beta} 
\left(1-\frac{m_W^2}{M_a^2}\right)
\left(1+\frac{2m_W^2}{M_a^2}\right), 
\label{NRdecayW}
\\
\Gamma(N_{R a}\to \overline{\nu_L^{}}_i Z)
&= \frac{\left|(U_\text{MNS}^\dag Y_\nu)_{i a}\right|^2 M_a}{64\pi \tan^2\beta} 
\left(1-\frac{m_Z^2}{M_a^2}\right)
\left(1+\frac{2m_Z^2}{M_a^2}\right), 
\label{NRdecayZ}
\end{align}
and also into neutrinos and scalar bosons ($h,H,A$) if kinematically allowed. 
Due to the strong suppression, i.e., $1/\tan^2\beta$, right-handed 
neutrinos may be long-lived. 
It seems difficult to consider the dark matter candidate, 
since the lifetime of right-handed neutrinos cannot be large 
compared to that of universe. 

In the third panel of FIG.~\ref{FIG:width}, 
 the total decay width of charged Higgs bosons is shown. 
For smaller values of $\tan\beta$, charged Higgs boson can decay 
 into quarks, while for large $\tan\beta$ they do into leptons. 
In the last panel of FIG.~\ref{FIG:width}, 
 the total decay width of right-handed neutrinos is given 
 as a function of $\tan\beta$. 
Because of the seesaw relation in Eq.\eqref{casasibara}, 
 $Y_\nu$ is proportional to $\tan\beta$, which cancels $\tan\beta$
 dependence in Eqs.\eqref{NRdecayW} and \eqref{NRdecayZ}. 
We found that the lifetime can be as large as $b$ quark, 
 so that secondary vertices may be tagged at the LHC. 
This property would be the same as $N_R^{}$ decay in the $B-L$ model 
(see, \cite{Ref:B-L}). 

\begin{figure}[tb]
\centering
\includegraphics[height=4cm]{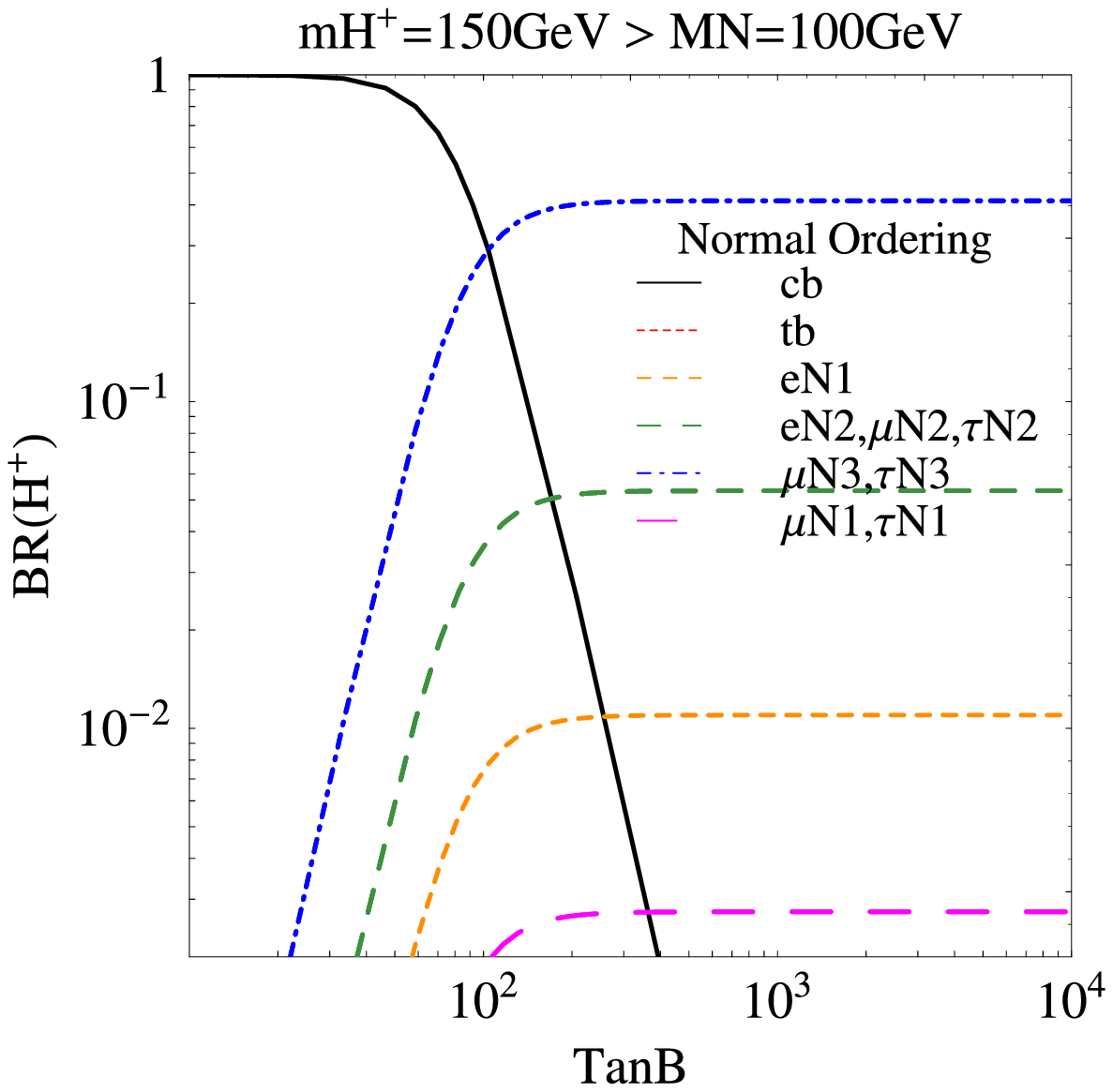}
\includegraphics[height=4cm]{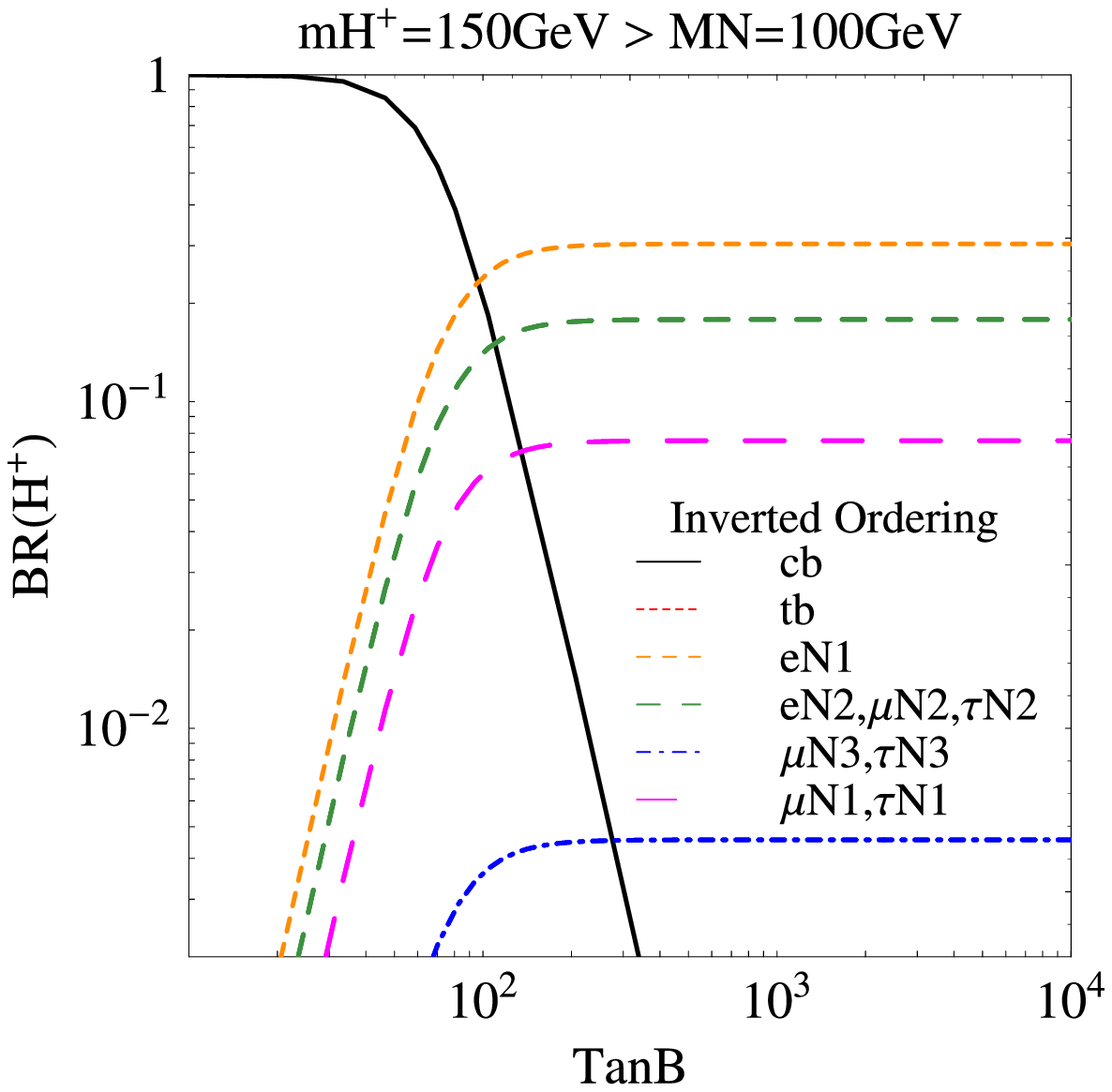}
\includegraphics[height=4cm]{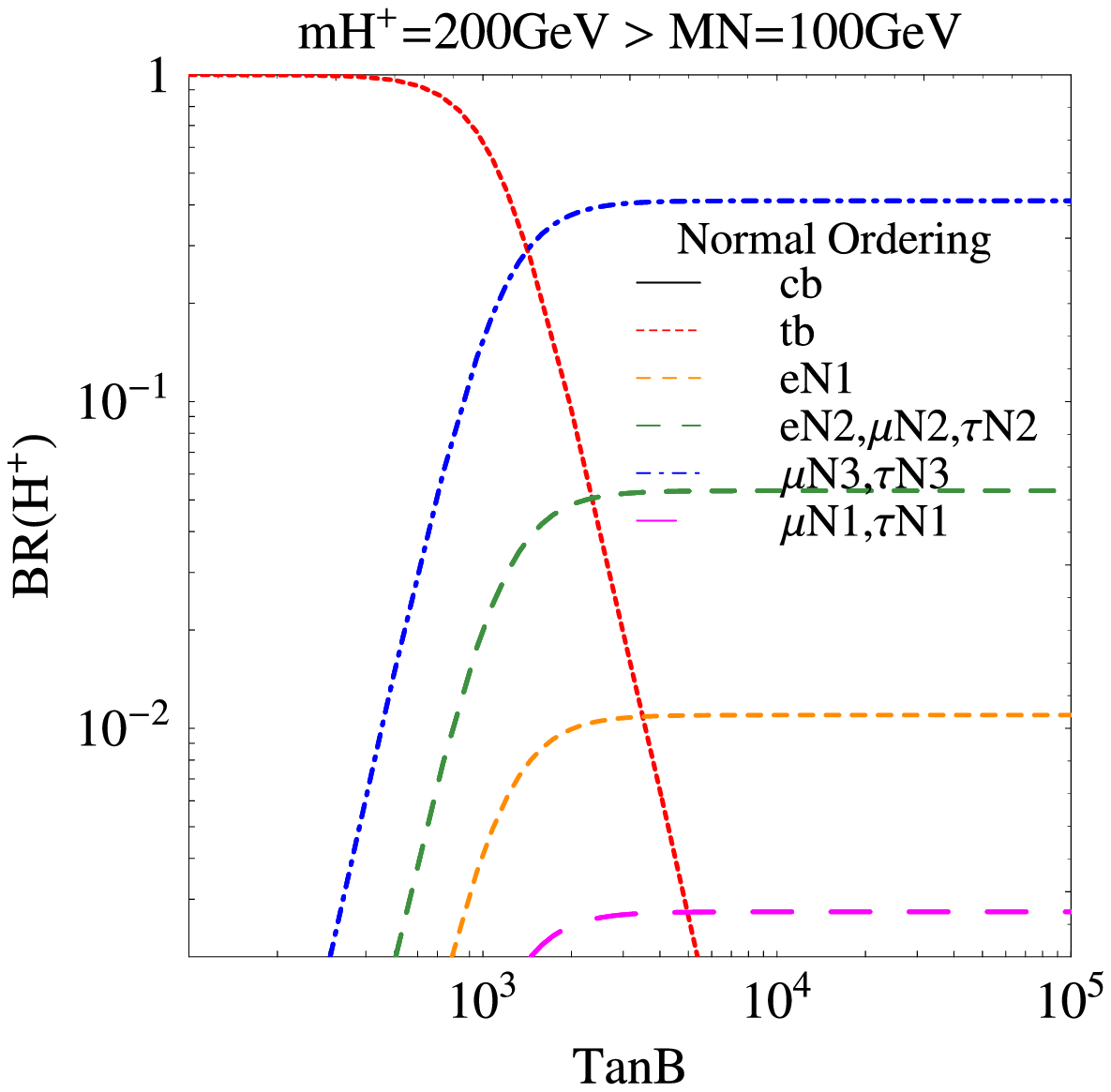}
\includegraphics[height=4cm]{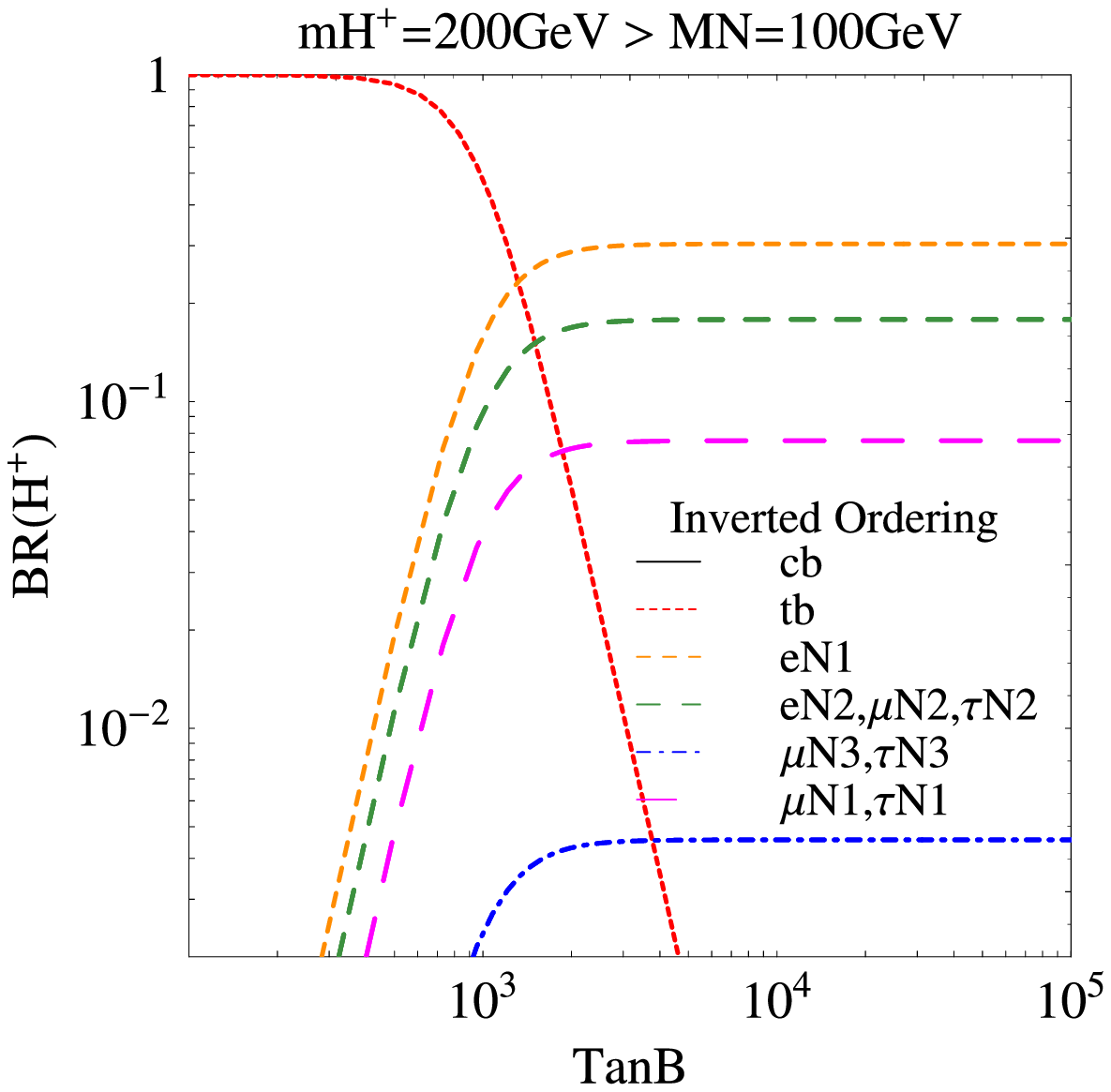}
\caption{Decay branching ratios of the charged Higgs boson are given for $m_{H^\pm}^{} > M_N$, 
where $R={\bf 1}$ and $M_N \propto {\bf 1}$ are taken.
}
\label{FIG:BrChA}
\end{figure}
In FIG.\ref{FIG:BrChA}, 
 we show decay branching fractions of charged Higgs boson,  
 where $m_{H^\pm}^{}=150$ GeV (first and second)
 and $200$ GeV (third and fourth) are taken. 
For lower $\tan\beta$ region, charged Higgs bosons mainly decay 
 into quark pairs, while for larger $\tan\beta$ region they do into 
 charged lepton and right-handed neutrino pairs. 
In large $\tan\beta$ region, main leptonic decay channels are 
 different from the case with $M_N > m_{H^\pm}^{}$.

\subsection{ILC phenomenology}

\begin{figure*}[tb]
\centering
\includegraphics[height=3.5cm]{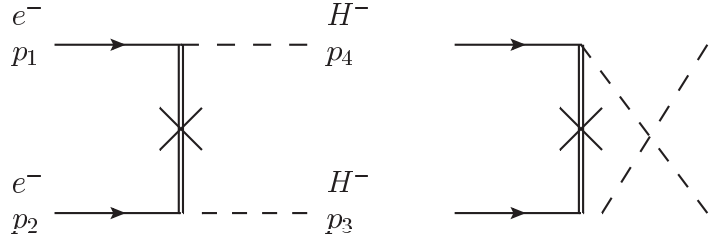}
\caption{The Feynman diagrams for $e^-e^-\to H^-H^-$ in $\nu$THDM with $N_R$. 
}
\label{FIG:eeHH}
\end{figure*}

Here let us investigate some ILC phenomenology. 
At first we focus on 
 an electron-electron collider, which is a possible option of the ILC. 
This would be the most important collision process for the $\nu$THDM to 
 be observed at the ILC. 
It is because, 
 in the $\nu$THDM,
 $e^-e^-$ collision can produce a pair of same sign charged Higgs bosons,
 where the lepton number violation is occurred in the Majorana mass term 
 of right-handed neutrinos. 
We stress that the discovery of this 
 lepton number violation is a very important key 
 of the beyond the SM. 
In FIG.\ \ref{FIG:eeHH}, Feynman diagrams for this process are depicted.  
The differential cross section is calculated as 
\begin{align}
\frac{d\sigma}{d\cos\theta}
&
= \frac{\beta_{H^\pm}}{256\pi}\left|Y_\nu \left(\frac{M_N}{\hat{t}-M_N^2}
+\frac{M_N}{\hat{u}-M_N^2}\right) Y_\nu^T\right|_{ee}^2
\simeq \frac{G_F^2 \beta_{H^\pm}}{8\pi} \tan^4\beta
\left| \langle M_\nu \rangle_{ee} \right|^2,
\end{align}
where $\theta$ is a scattering angle, and 
${\hat t}$ and ${\hat u}$ are the Mandelstam variables. 
The effective mass for the neutrinoless double beta decay is defined as 
$\langle M_\nu \rangle_{ee} = U_\text{MNS} m_\nu^\text{diag} U^T_\text{MNS}$. 
In the last step, we use the seesaw relation and require ${\hat t},{\hat u} \ll M_N^2$ 
 to reduce the formula. 

\begin{figure*}[tb]
\centering
\includegraphics[height=4.5cm]{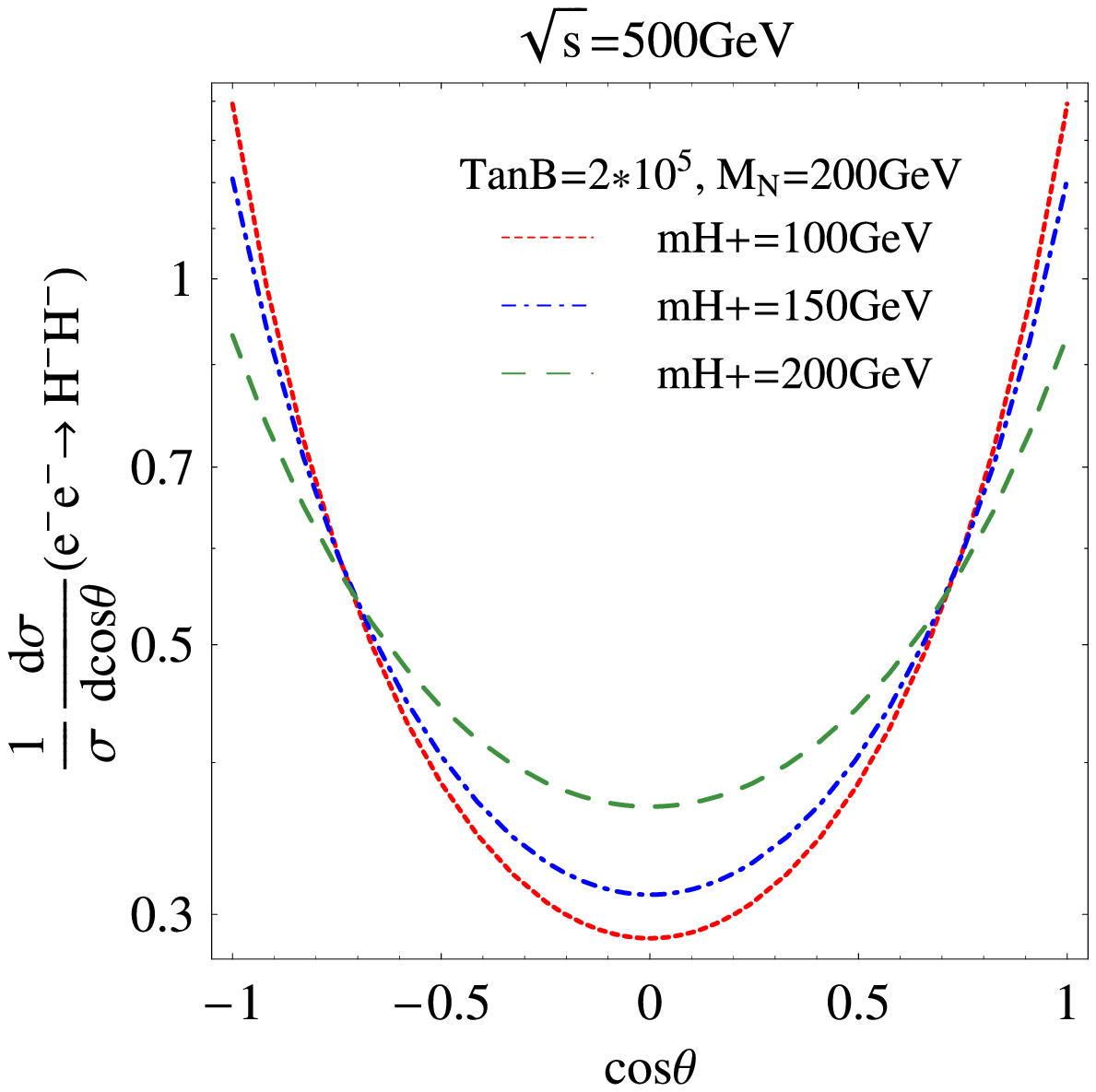}
\includegraphics[height=4.5cm]{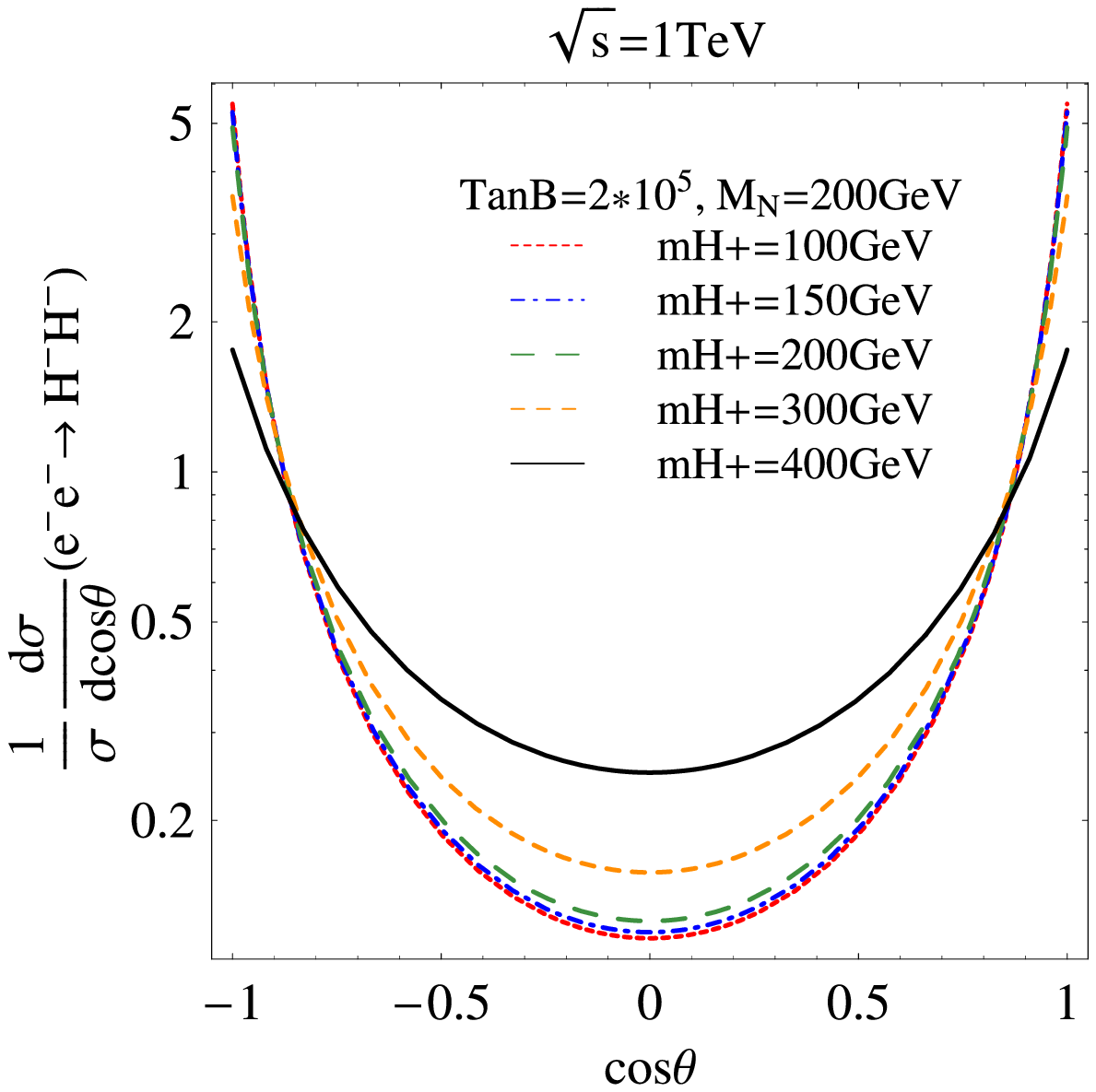}
\caption{Differential cross sections of $e^-e^-\to H^-H^-$ in $\nu$THDM with $N_R$. 
Mass of right-handed neutrinos is set as $M_N=200$ GeV. 
}
\label{FIG:eeHHdxdcMN200}
\end{figure*}
Differential cross sections 
 of $e^-e^-\to H^-H^-$ with fixed $M_N$ are shown in FIGs.~\ref{FIG:eeHHdxdcMN200}, 
 where $R={\bf 1}$ and $M_N \propto {\bf 1}$ are taken. 
Collision energies are chosen as $500$ GeV and 
 $1000$ GeV, respectively. Since we take $M_N \propto {\bf 1}$, 
 the normal ordering case and inverted ordering case of the light neutrino mass spectrum 
 give the same distribution. 
The ratio of VEVs are set to be $\tan\beta= 2\times 10^5$, 
 which is allowed by the stringent constraint from $\mu \to e\gamma$. 
We can easily see that scattered charged Higgs bosons mainly go to the forward direction. 
This is a typical behaviour of the co-linear enhancement of the cross section due to 
 the $t(u)$-channel diagram. 
\begin{figure*}[tb]
\centering
\includegraphics[height=4.1cm]{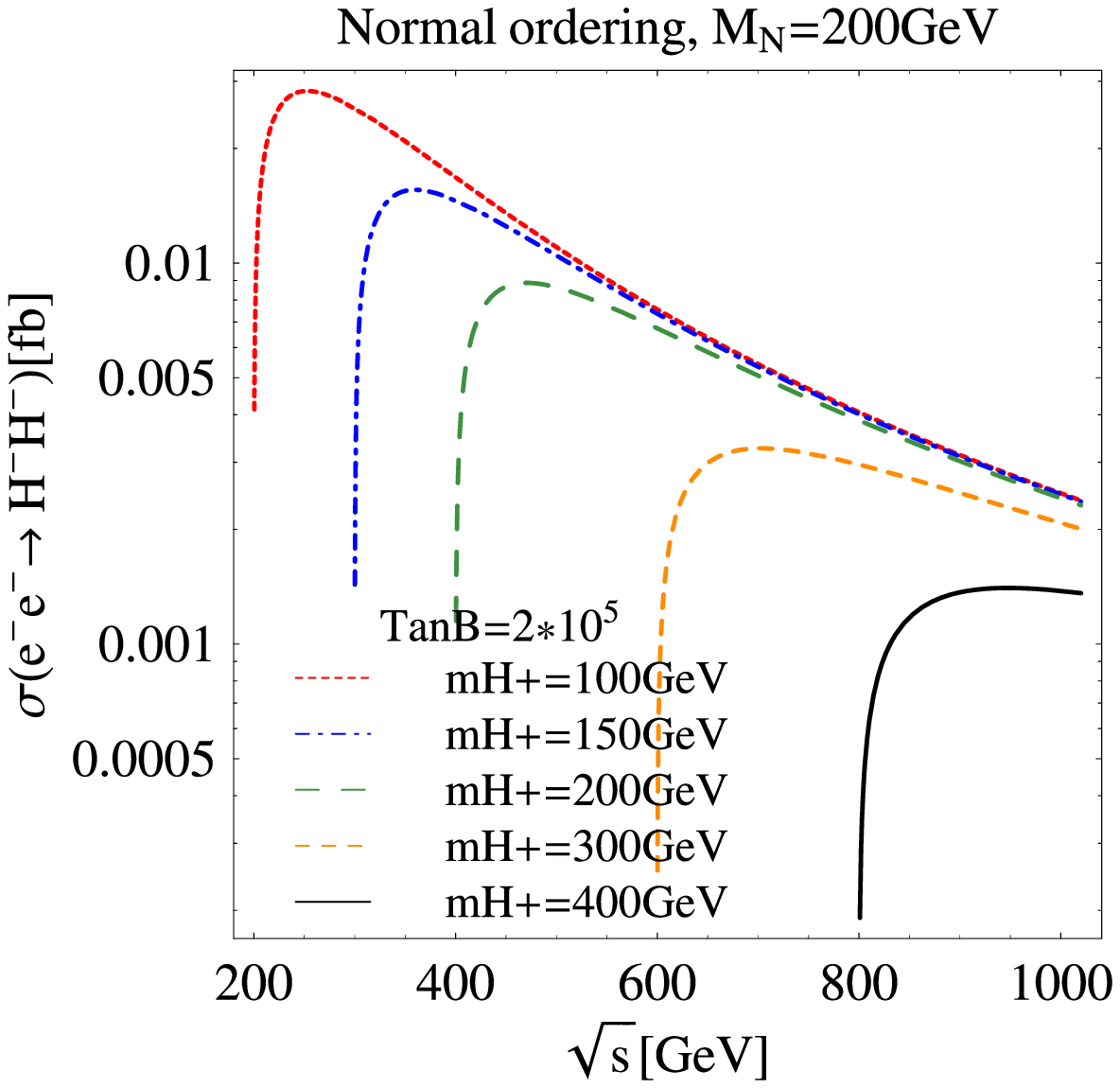}
\includegraphics[height=4.1cm]{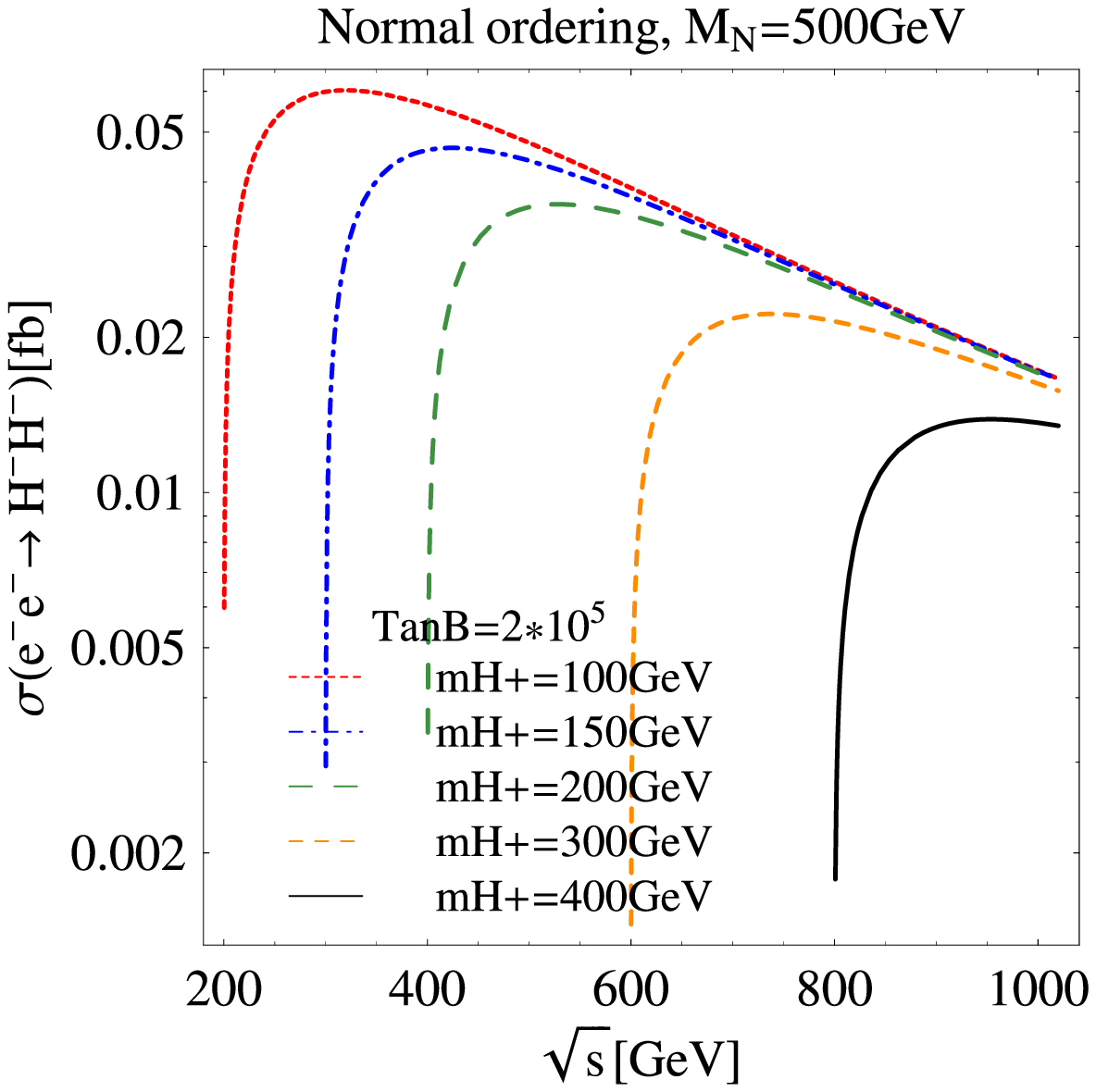}
\includegraphics[height=4.1cm]{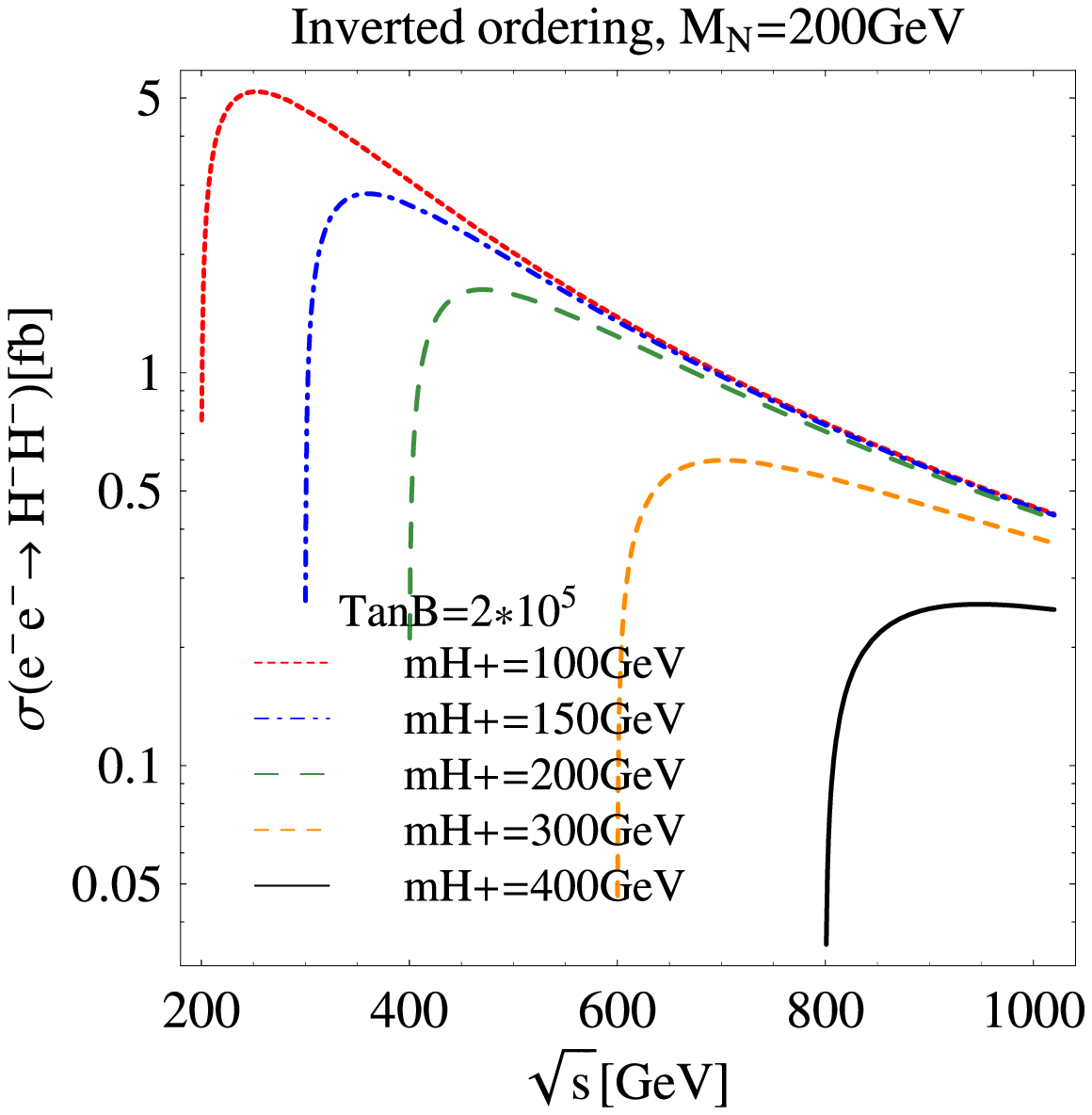}
\includegraphics[height=4.1cm]{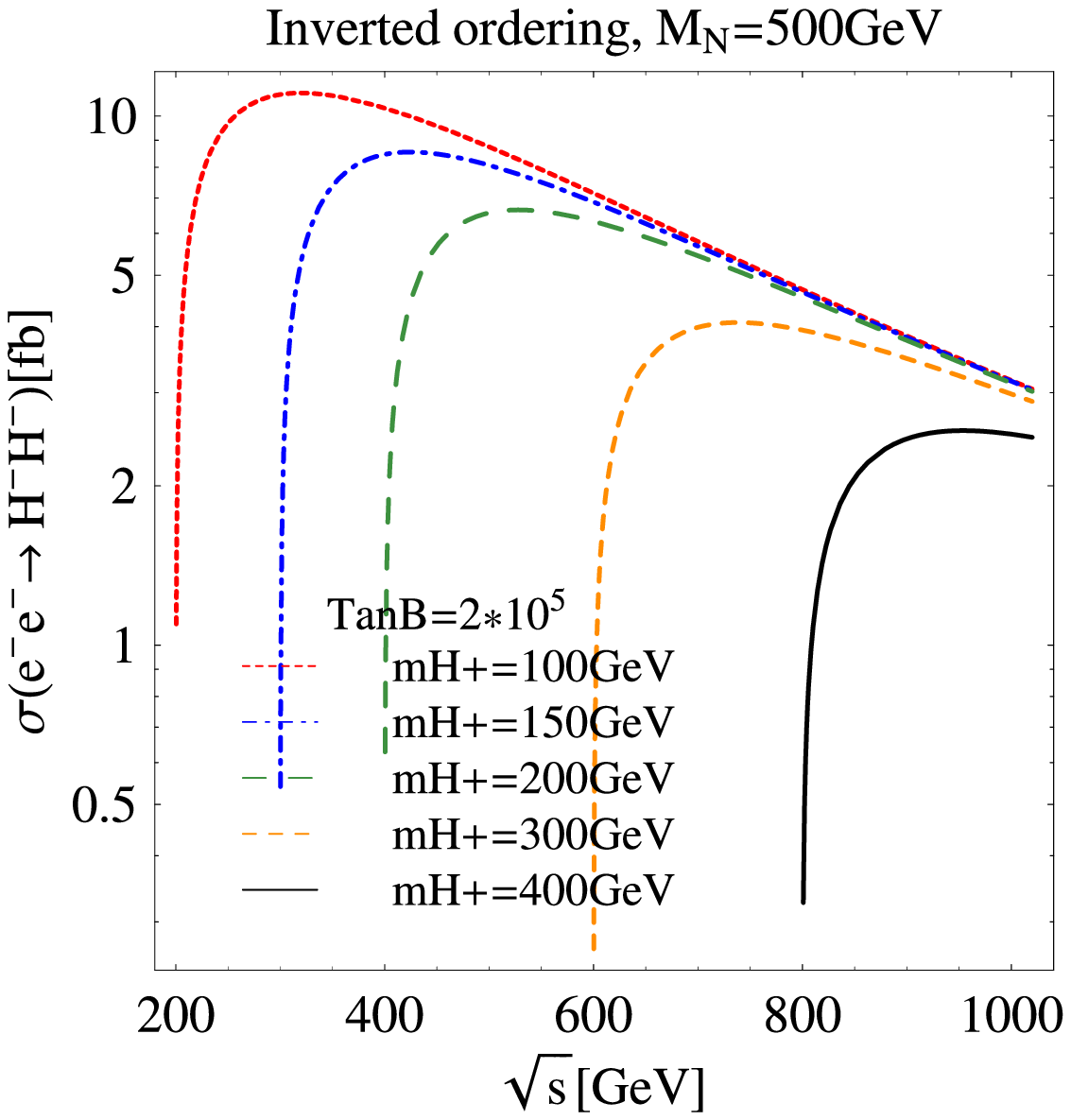}
\caption{Total cross sections of $e^-e^-\to H^-H^-$ in $\nu$THDM with $N_R$. 
Mass of right-handed neutrinos is set as $M_N=200$ GeV (first and second) and 
 $500$ GeV (third and fourth).
}
\label{FIG:eeHHxMN}
\end{figure*}
Performing angle integration, total cross sections are evaluated as a function of 
 the collision energy in FIGs.~\ref{FIG:eeHHxMN}. 
Masses of right-handed neutrinos are taken to be $200$ GeV and $500$ GeV, respectively.
Since we used fixed value of $\tan\beta$, heavier right-handed neutrinos give a larger 
 cross section. 
Reflecting the neutrino Yukawa matrix, we found that total cross sections can be 
 as large as $0.01$ fb ($1$fb) for the normal (inverted) ordering of the light 
 neutrino mass spectrum, if $R={\bf 1}$ and $M_N \propto 1$ are satisfied. 
Polarized beams would be useful to explore the structure of the interaction vertex.
Because the vertex is originated from the Yukawa interaction, use of left-handed 
 electrons can enhance cross sections. 

\begin{figure*}[tb]
\centering
\includegraphics[height=4.5cm]{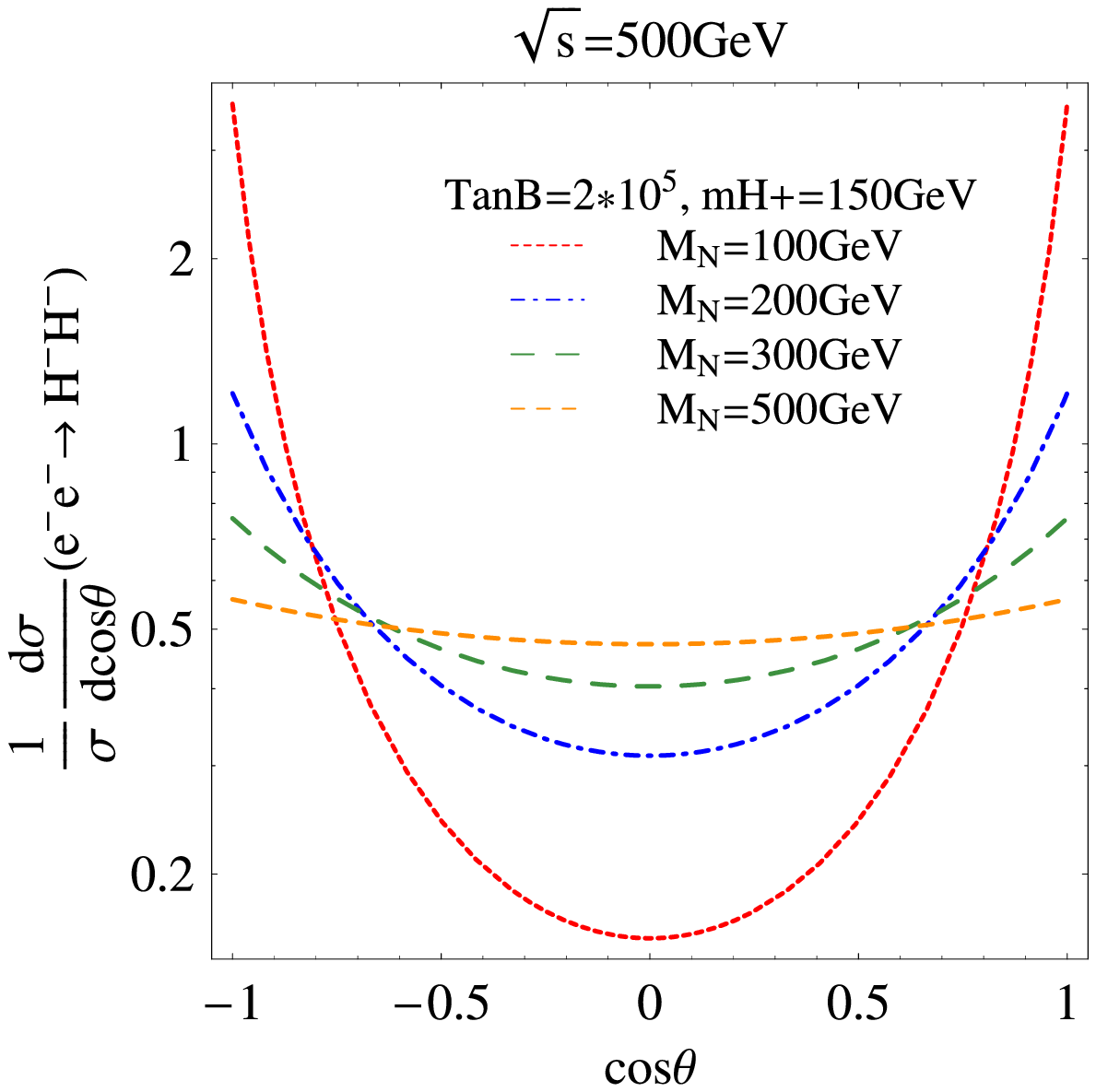}
\includegraphics[height=4.5cm]{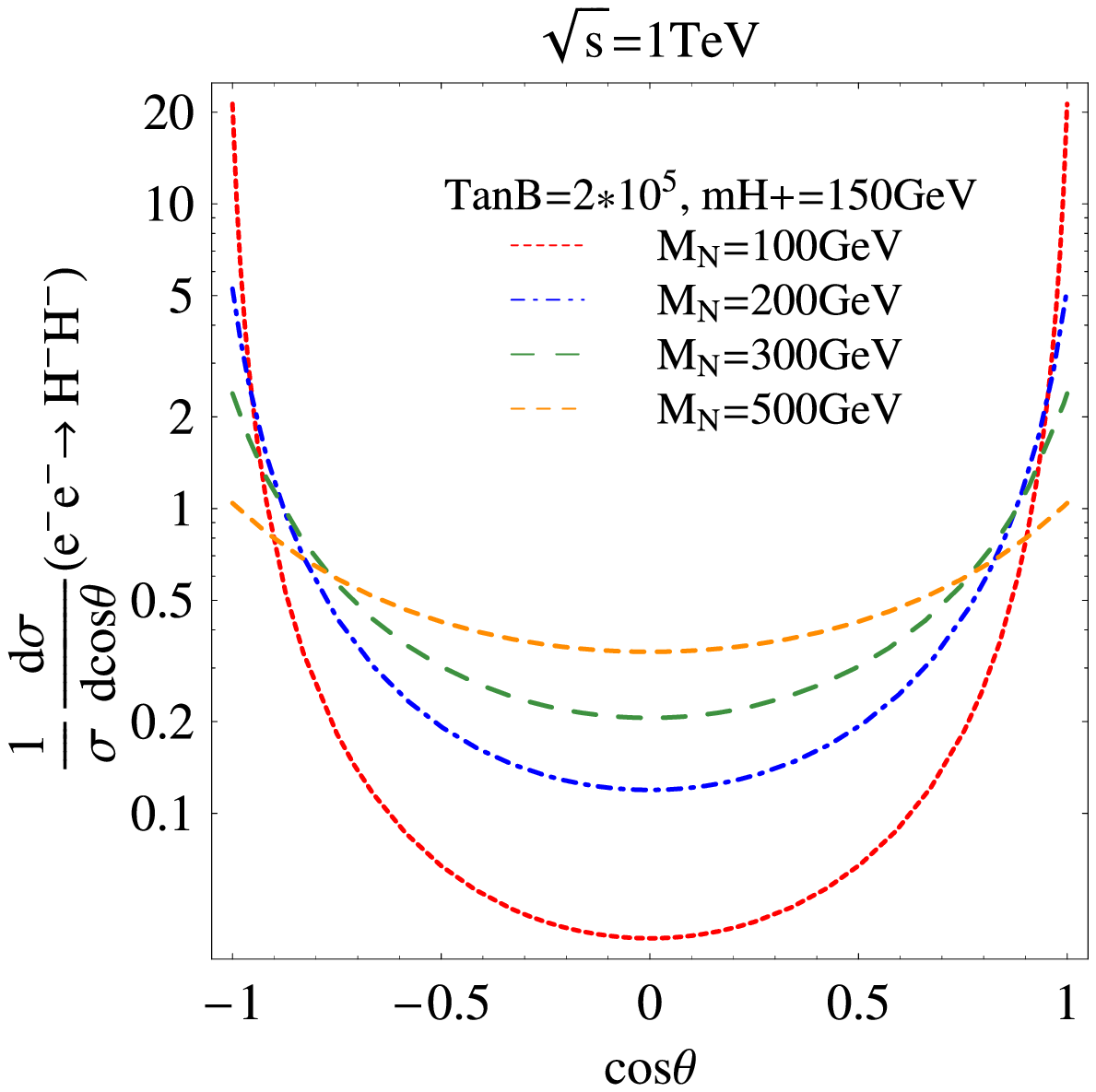}
\caption{Differential cross sections of $e^-e^-\to H^-H^-$ in $\nu$THDM with $N_R$. 
Mass of charged Higgs bosons is taken to be $m_{H^\pm}=150$ GeV. 
}
\label{FIG:eeHHdxdcmH150}
\end{figure*}
In FIG.~\ref{FIG:eeHHdxdcmH150}, we show angular distributions of $e^-e^-\to H^-H^-$ 
 cross sections for $m_{H^\pm}=150$ GeV. 
Masses of right-handed neutrinos are taken as $100, 200, 300$ and $500$ GeV.
As we mentioned before, cross sections become larger for heavier masses of 
 right-handed neutrinos.
\begin{figure*}[tb]
\centering
\includegraphics[height=4.1cm]{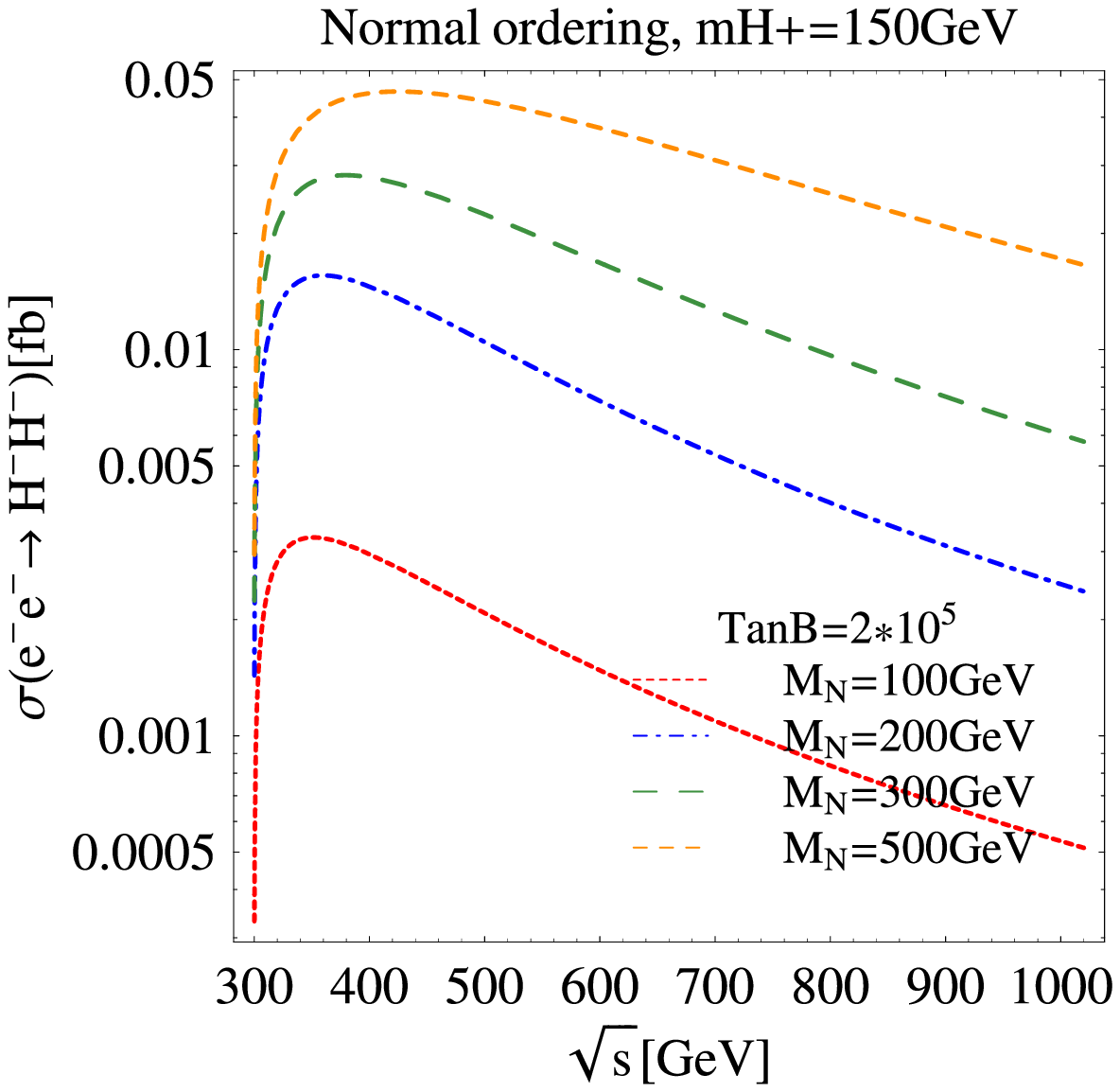}
\includegraphics[height=4.1cm]{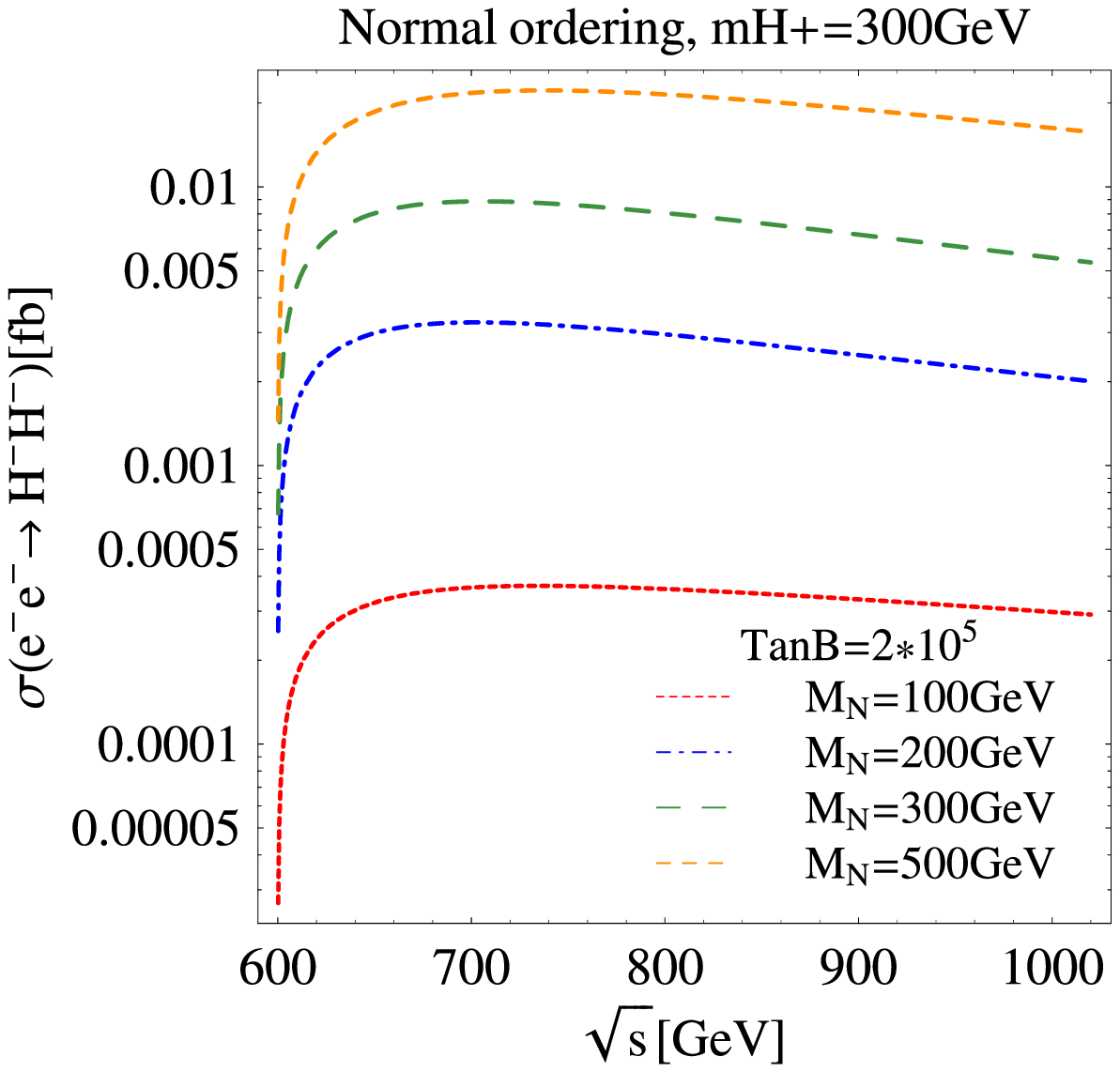}
\includegraphics[height=4.1cm]{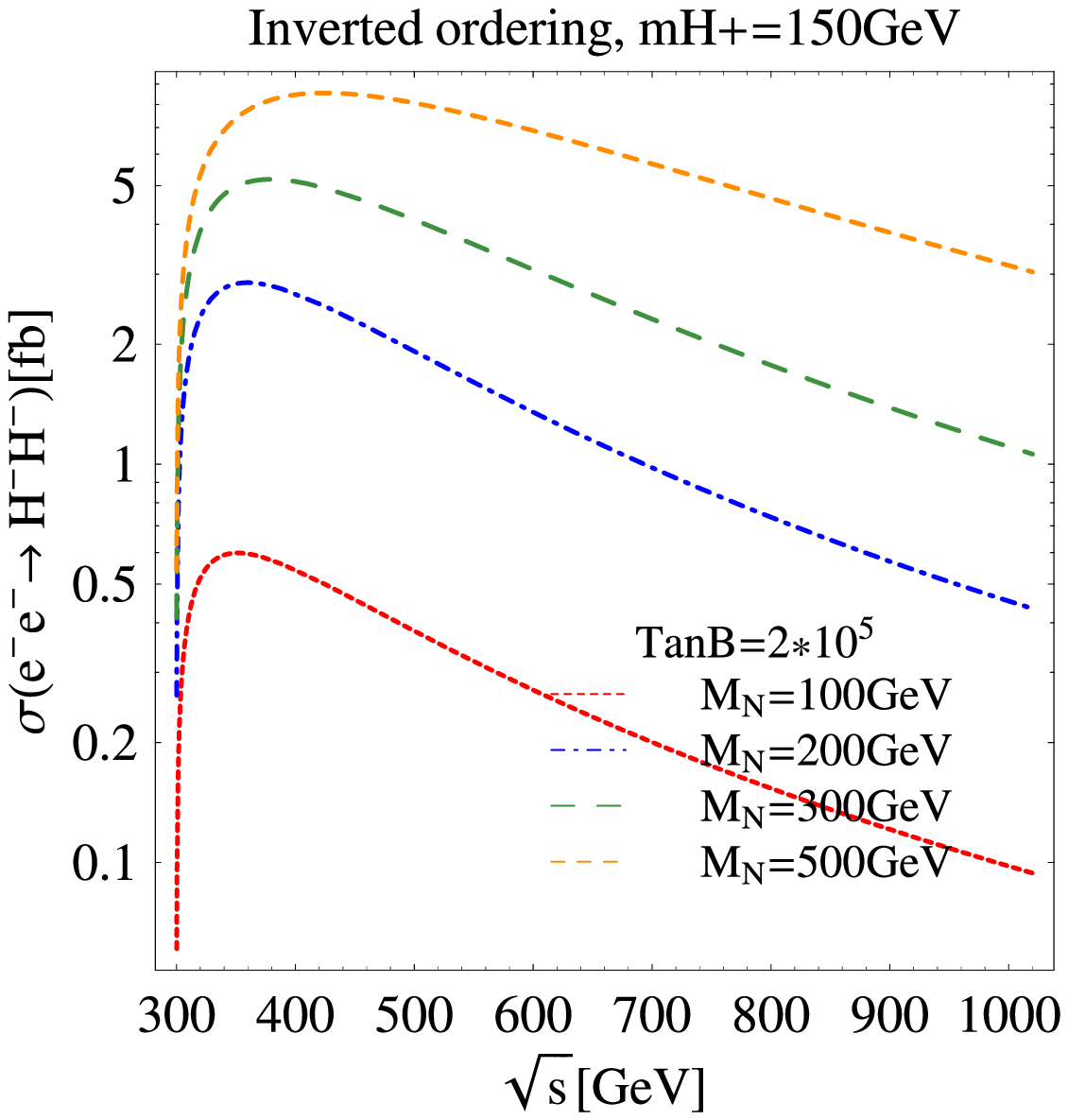}
\includegraphics[height=4.1cm]{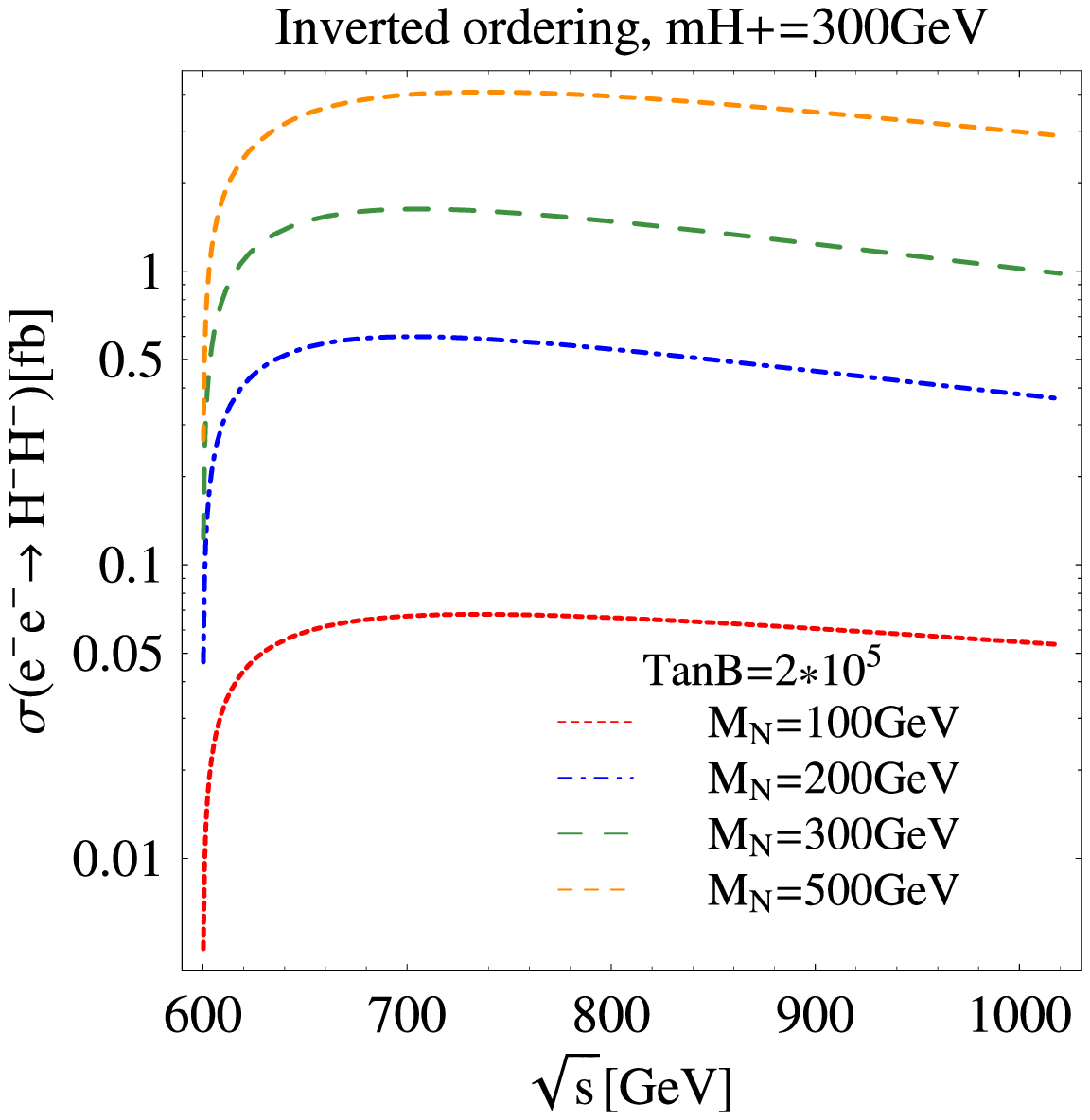}
\caption{Total cross sections of $e^-e^-\to H^-H^-$ in $\nu$THDM with $N_R$. 
Mass of charged Higgs bosons is taken to be $m_H=150$ GeV and $300$ GeV.
}
\label{FIG:eeHHxmH}
\end{figure*}
In FIG.~\ref{FIG:eeHHxmH}, total cross sections are shown as a function of 
 collision energies by changing the mass of right-handed neutrinos. 

Next, we consider the usual $e^+e^-$ collision, 
 the charged Higgs boson pair production via 
 the Drell-Yann process $e^+e^-\to H^+H^-$. 
Since main decay modes of $H$ and $A$ can be neutrinos, 
$e^+e^-\to A H$ process is also useful as for a source of $N_R$.
Production cross sections are evaluated in Refs.\cite{DYChee}.
In our model, charged Higgs boson can be light, 
 significant number of charged Higgs boson pairs 
 can easily be created, and its specific 
 leptonic decay mode would be detectable. 
Furthermore, $t$-channel exchange of right-handed neutrinos can 
 also contribute to the process. 
Because of the chiral structure of the Yukawa interaction, 
 only $\sigma^{LL}$ is changed in the massless limit of electrons, 
 where $LL$ denotes the initial electron and positron polarizations.
Differential cross sections are given by
\begin{align}
\frac{d\sigma^{LL}_{}}{d{\hat t}} = \frac{d\sigma^{LL}_0}{d{\hat t}} \left[ 1 
+ \left(\frac1{8\pi\alpha} Y_\nu \frac{\hat s}{{\hat t}-M_N} Y_\nu^\dag \right)_{ee} \right]^2, 
\end{align}
where ${\hat s}$ and ${\hat t}$ are the Mandelstam variables, 
 and $\sigma_0$ is the production cross section of $e^+e^-\to H^+H^-$ 
 in the THDM without right-handed neutrinos.
The impact of the second term can be one percent level, 
 which may be detectable in the precise measurement of 
 the left-right polarization asymmetry.

The last is about $e^-\gamma$ collision, 
 which is another option of the ILC.  
Relatively narrow band photons can be generated from electrons and laser beams 
 by the compton back-scattering method. 
In this option, $e^-\gamma \to H^-N$ process may be promising to produce 
 right-handed neutrinos. 
Production cross sections are not suppressed so strongly 
 because cross sections are only proportional to $Y_\nu^2$\cite{eGam}.

Finally,
 we comment on the production process of right-handed neutrinos at the ILC. 
Although heavy right-handed neutrinos are gauge singlet, 
 which can interact with electron at tree level via the neutrino Yukawa interaction. 
The Yukawa couplings can be sizable even under the stringent constraint from 
 the $\mu\to e\gamma$. 
Thus, right-handed neutrinos can be produced by the $t$-channel exchange of 
 charged Higgs bosons at ILC as $e^+e^-\to N\overline{N}$. 
The cross sections would be similar to $e^-e^-\to H^-H^-$, 
 because production rates are proportional to $Y_\nu^4$. 
Since right-handed neutrinos only couple to the left-handed electron 
 in the massless limit, the production cross section can also be controlled 
 by initial electron and positron polarizations, which may make precision 
 measurement possible.

\section{Summary and discussions}

Smallness of neutrino mass is explained 
 by tiny VEV of 
 extra-Higgs doublet which couples to 
 neutrinos. 
This model is so-called $\nu$THDM where 
 TeV-scale seesaw works well, 
 and could be 
 observable at LHC and ILC experiments. 
Notice that
 neutrino Yukawa couplings are not tiny in the $\nu$THDM,
 which makes the model tested at LHC or ILC. 
We have investigated collider signatures of
 the $\nu$THDM, 
 and shown characteristic signals 
 can be observable at LHC. 
For example, 
 detective charged tracks can be obtained from  
 long lived charged Higgs 
 when $m_{H^\pm}^{} < M_N$. 
On the other hand, 
 when $m_{H^\pm}^{} > M_N$, 
 the charged Higgs bosons mainly decay into 
 a lepton and a right-handed neutrinos.  
Then right-handed neutrinos can be long-lived 
 similarly to the $B-L$ model, and 
 secondary vertices may be tagged at the LHC. 
We have also investigated ILC phenomenology such as
 lepton number violating processes at $e^-e^-$ collision.
The discovery of this  
 lepton number violation
 is a very important key of the beyond the SM.


\vspace{1cm}

{\large \bf Acknowledgments}\\

\noindent
We thank O. Seto, S. Matsumoto, S. Kanemura, M. Aoki and H. Sugiyama
 for useful and helpful discussions. 
This work is partially supported by Scientific Grant by Ministry of 
 Education and Science, Nos. 20540272, 20039006, and 20025004.


\end{document}